\documentclass{aa}
\usepackage{times}
\usepackage[dvips]{epsfig}
\def\bbbn{{\rm I\!N}}

\begin{document}
\thesaurus{05(12.07.1,08.02.3,08.16.2,10.19.2)}
\title{The binary gravitational lens and its extreme cases}
\author{M. Dominik\inst{1,2}\thanks{Work carried out at the Space
Telescope Science Institute has been financed by a research grant from
Deutsche Forschungsgemeinschaft (Do 629-1/1) and work carried out at the Kapteyn
Astronomical Institute is financed by a Marie Curie Fellowship from the
European Union (FMBICT972457). {\em e-mail:} {\tt dominik@astro.rug.nl}}}
\institute{Space Telescope Science Institute, 3700 San Martin Drive, Baltimore,
MD 21218, USA \and Kapteyn Astronomical Institute, Postbus 800, NL-9700 AV Groningen,
The Netherlands}
\date{Received ; accepted}
\maketitle
\begin{abstract}
The transition of the binary gravitational lens from the equal mass case to
small (planetary) mass ratios $q$ is studied. It is shown how the 
limit of a (pure shear) Chang-Refsdal lens is approached, under what conditions the Chang-Refsdal approximation
is valid, and how the 3 different topologies of the critical curves and caustics for a 
binary lens are mapped onto the 2 different topologies for a Chang-Refsdal lens with pure 
shear. It is shown that for wide binaries, the lensing in the
vicinity of both lens objects can be described by 
a Taylor-expansion of the deflection term due to the other object, where the Chang-Refsdal approximation
corresponds to a truncation of this series. For close binaries, only the vicinity of the secondary,
less massive, object can be described in this way. 
However, for image distances much larger than the separation of the lens objects,
any binary lens can be approximated by means
of multipole expansion, where the
first non-trivial term is the quadrupole term. 
It is shown that an ambiguity exists between wide and close binary lenses, 
where the shear at one of the objects due
to the other object for the wide binary 
is equal to the absolute value of the eigenvalues of the quadrupole moment for the close
binary.
This analysis provides the basis for a classification of binary lens microlensing events, especially
of planetary events, and an understanding of present ambiguities. 
\end{abstract}


\section{Introduction}
The binary gravitational lens is the appropriate model with which to describe lensing by 
a binary star or by a star surrounded by a planet. While both systems fit
into this model, there are some typical properties of the binary lens that 
essentially differ in both cases. The aim of this paper is to study the
properties of binary lenses in the limit of extreme mass ratio and  
extremely close and wide separations,
and to show explicitly how these limits arise and under what conditions the
approximation of the binary lens by the simplified limiting cases is successful or not.

The binary lens for the equal mass case has been discussed in great detail
by Schneider \& Wei{\ss} (\cite{SchneiWei}), who show 
that, depending on the distance between the two lens objects, there are
3 different topologies of the critical curves and caustics.
Erdl \& Schneider (\cite{Erdl}) have shown that the same 3 topologies of the critical
curves and caustics apply to an arbitrary mass ratio and have determined the
separations between the lens objects for which transitions between these 3 topologies
occur.

On the other side, Chang \& Refsdal (\cite{CRlens}, \cite{CR2}) have discussed a lens model,
now known as the 'Chang-Refsdal lens' in which
an effective distortion term,
characterized by the 2 parameters convergence $\kappa$ and
shear $\gamma$, 
is added to a point-mass lens. 
Their aim was to 
discuss the effect of a star in a lensing galaxy yielding to a splitting of one of the
images due to the galaxy into microimages and thereby altering the observed image
flux. This effect is the original meaning of 'microlensing'.

The relationship between the Chang-Refsdal lens and the binary lens has first been 
discussed by Schneider \& Wei{\ss} (\cite{SchneiWei}) yielding a relation between
the shear $\gamma$ and the lens separation $d$ and the mass ratio $q$ for wide
binary systems.
This Chang-Refsdal limit has been rediscovered in the discussion of the binary lens
models for the MACHO LMC-1 event (Dominik \& Hirshfeld~\cite{DoHi2}).

It has been mentioned that the effect of a planet around a lens star can
be described as lensing by the star and a planetary distortion involving 
a Chang-Refsdal lens (Gould \& Loeb~\cite{GL};
Gaudi \& Gould~\cite{GG}).

Griest \& Safizadeh (\cite{GS}) have discussed the case in which 
the source star passes close to the primary lens star and therefore exhibits 
a high peak magnification. In this case, a small 'central caustic' is located
near the primary lens star yielding deviations from the point-lens case.
Gaudi et al.~(\cite{GNS}) have pointed out that this central caustic is
affected by all planets around the lens star rather than by a single planet.

Although there have been discussions of some binary lens limits, there has
not yet been a comprehensive and systematic discussion of all of these effects.
This discussion combines the different views on specific subcases discussed in
previous papers and puts them into a new context, yielding 
new interesting conclusions and a comprehensive classification.
In particular, the transition towards the extreme cases and the validity of the
limits as approximations is poorly discussed in the literature.
I present expansions that yield the exact result as infinite series 
and approximations as finite series.

The applications of the results derived in this paper for the characterization of binary lens
microlensing events and their ambiguities will be discussed in a separate paper
(Dominik \& Covone, in preparation).

In Sect.~\ref{lenseq}, the basic quantities relevant to the discussion
are defined.
In order to study how the binary lens approaches its limiting cases, 
it is 
necessary to review some properties of the Chang-Refsdal lens in Sect.~{\ref{sec:CR}, as
well as to discuss some basic properties of the quadrupole lens in
Sect.~\ref{sec:quad}. 
In Sect.~\ref{sec:bintop}, the 3 different topologies of the critical curves and
caustics of the binary lens as a function of mass ratio are discussed. 
The vicinity of the secondary object is investigated in Sect.~\ref{sec:vicsec} by means of the full
lens equation and a Taylor-expansion of the deflection due to the primary object. It is shown how the
Chang-Refsdal caustics evolve from the binary lens caustics as $q \to 0$.
In Sect.~\ref{sec:perturbation}, the results of Sect.~\ref{sec:vicsec} are compared with the perturbative
picture of Gould \& Loeb (\cite{GL}) and Chang \& Refsdal (\cite{CRlens}).
In Sect.~\ref{sec:centralcaust}, the behaviour of the central caustic is discussed and it is shown
that there is an ambiguity between close and wide binaries, in which the shear near one object due to the
other object for the wide binary
is equal to the absolute value of the eigenvalues of the quadrupole moment for the close binary.
Finally, in Sect.~\ref{sec:sim}, the similarities and differences of diamond-shaped caustics produced
by binary lenses, quadrupole lenses, and pure shear Chang-Refsdal lenses are shown.
Two appendices show the derivation of the Taylor-expansion in real
notation
(Appendix~A) and in complex notation (Appendix B). In Appendix B, the convergence radius of the series
is also derived.

\section{The lens equation}
\label{lenseq}
Consider a source star at distance $D_\mathrm{S}$ from the observer, a
lens composed of $N$ point-like objects at distance 
$D_\mathrm{L}$ from the observer, and let the
distance between lens and source be $D_\mathrm{LS}$.{\footnote{The difference
in the distances of the lens objects along the line-of-sight is  
assumed to be much smaller
than $D_\mathrm{L}$ and $D_\mathrm{LS}$, and neglected.} With $M$ being the total mass of the
lens, the characteristic angular scale of this system for gravitational
lensing is the angular Einstein radius $\theta_\mathrm{E}$ given by
\begin{equation}
\theta_\mathrm{E} = \sqrt{\frac{4GM}{c^2}\,\frac{D_\mathrm{LS}}{
D_\mathrm{L}\,D_\mathrm{S}}}\,.
\end{equation}
The lens equation yields the true angular position of the source object
$\vec \beta$ as a function
of its apparent (lensed) angular position $\vec \theta$. 
With dimensionless coordinates
$\vec x = \vec \theta/\theta_\mathrm{E}$ and $\vec y
= \vec \beta/\theta_\mathrm{E}$, 
the lens equation reads
\begin{equation}
\vec y = \vec x - \vec \alpha(\vec x)\,,
\label{eq:lenseqgen}
\end{equation}
where $\vec \alpha(\vec x)$ gives the deflection due to the lens, and can
written as the gradient of the deflection potential $\psi$
(e.g.~Schneider et al.~\cite{SEF}, p.~159)
\begin{equation}
\vec \alpha(\vec x) = \vec \nabla \psi(\vec x)\,.
\label{eq:gradpot}
\end{equation}
Consider $N$ lens objects with mass fractions $m_r$ ($r = 1\ldots{}N$),
where 
\begin{equation}
\sum_{r=1}^{n} m_r = 1\,,
\end{equation}
located at angular positions $\vec x^{(r)}\,\theta_{\rm E}$. 
The deflection potential $\psi$ is then given by
\begin{equation}
\psi(\vec x) = \sum_{r=1}^{N} m_r\,\ln \left|\vec x - \vec x^{(r)}\right|\,,
\label{eq:psi}
\end{equation}
so that the lens equation reads
\begin{equation}
\vec y = \vec x - \sum_{r=1}^{N} m_r\,\frac{\vec x - \vec x^{(r)}}
{|\vec x - \vec x^{(r)}|^2}\,.
\label{eq:lenseq}
\end{equation}

An important characteristic of a given lens model are the positions
and shapes of the {\em critical curves} and the {\em caustics}.
The critical curves $\cal C_\mathrm{crit}$
are defined by the set of points in the space of 
observed angular positions for which the Jacobian determinant of
the lens mapping vanishes, i.e.
\begin{equation}
{\cal C}_\mathrm{crit} = \left\{\vec x_\mathrm{crit} | 
\det \left(\frac{\partial \vec y}{\partial \vec x}\right)
(\vec x_\mathrm{crit}) = 0\right\}\,,
\end{equation}
while the caustics $\cal C_\mathrm{caust}$ are formed by the set of 
points in true source position space onto which the critical curves are mapped, i.e.
\begin{eqnarray}
{\cal C}_\mathrm{caust} & = & \bigg\{\vec y_\mathrm{caust} | 
\vec y_\mathrm{caust} = \vec y(\vec x_\mathrm{crit}), \nonumber \\
& & \quad \quad \left.
\det \left(\frac{\partial \vec y}{\partial \vec x}\right)(\vec x_\mathrm{crit}) = 0\right\}\,.
\end{eqnarray}
As a source touches a caustic, two of its images merge at a critical curve
and disappear once the caustic has been crossed. This means that the number
of images for a given source changes by two as a source crosses a 
caustic (e.g.\ Chang \& Refsdal~\cite{CRlens}).

For a binary lens, consider object 1 with mass fraction $m_1$ being placed at the angular position
$(d/2,0)\,\theta_\mathrm{E}$ 
and object 2 with mass fraction
$m_2 = 1-m_1$ at the angular position $(-d/2,0)\,\theta_\mathrm{E}$,
so that $d$ is the angular separation between the lens objects in units of $\theta_\mathrm{E}$.
From Eq.~(\ref{eq:lenseq}), the lens equation then reads
\begin{eqnarray}
y_1 (x_1,x_2) & = & x_1 - m_1\,\frac{x_1-d/2}{(x_1-d/2)^2 + x_2^2} - \nonumber \\
& & \quad - (1-m_1)\,\frac{x_1+d/2}
{(x_1+d/2)^2+x_2^2}\,, \nonumber \\ 
y_2 (x_1,x_2) & = & x_2 - m_1\,\frac{x_2}{(x_1-d/2)^2 + x_2^2} - \nonumber \\
& & \quad - (1-m_1)\,\frac{x_2}
{(x_1+d/2)^2+x_2^2}\,.  
\label{eq:label7}
\end{eqnarray}

In some cases, is more convenient
to write the lens equation in different ways, e.g. if one wants to study objects of very different masses (e.g.
a star and a planet).
Throughout this paper,  object 1 is chosen to be the heavier, 'primary', object,
i.e. $m_1 \geq 0.5$, while object 2 will be the less massive, 'secondary', object.
For planetary events, the primary object is the star
and the secondary object is the planet, but to avoid being restricted to this
astrophysical application, I will not talk about 'star' and 'planet'.

Let us define the mass ratio between the secondary and the primary object as
\begin{equation}
q = \frac{m_2}{m_1} = \frac{1-m_1}{m_1}\,,\label{eq:defq}
\end{equation}
where $0 < q \leq 1$. The 
angular Einstein radii of the individual objects
$\theta_\mathrm{E}^{\,[1]}$ and
$\theta_\mathrm{E}^{\,[2]}$ are given by  
\begin{eqnarray}
\theta_\mathrm{E}^{\,[1]} & = & \theta_\mathrm{E}\,\sqrt{m_1}
 = \frac{\theta_\mathrm{E}}{\sqrt{1+q}}\,, \nonumber \\
\theta_\mathrm{E}^{\,[2]} & = & \theta_\mathrm{E}\,\sqrt{1-m_1}
 = \theta_\mathrm{E}\sqrt{\frac{q}{1+q}}\,,
\end{eqnarray}
and the angular separations $d^{\,[1]}$ and $d^{\,[2]}$ between the lens objects in
units of
$\theta_\mathrm{E}^{\,[1]}$ and
$\theta_\mathrm{E}^{\,[2]}$ read  
\begin{eqnarray}
d^{\,[1]} & = & \frac{d}{\sqrt{m_1}} = d\,\sqrt{1+q}\,, \nonumber \\
d^{\,[2]} & = & \frac{d}{\sqrt{1-m_1}} = d\,\sqrt{1+q^{-1}} = \frac{d^{\,[1]}}{\sqrt{q}}\,.
\label{eq:d1d2def}
\end{eqnarray}

By defining coordinates $\vec x^{\,[1]}$ and $\vec y^{\,[1]}$ centered on the primary
object that denote angular positions in units of $\theta_\mathrm{E}^{\,[1]}$,
the lens equation reads
\begin{eqnarray}
y_1^{\,[1]}(x_1^{\,[1]},x_2^{\,[1]}) & = & x_1^{\,[1]} - 
\frac{x_1^{\,[1]}}{\left[x_1^{\,[1]}\right]^2+\left[x_2^{\,[1]}\right]^2} - \nonumber \\
& & \quad 
- \,q\,\frac{x_1^{\,[1]} - d^{\,[1]}}{\left(x_1^{\,[1]} -d^{\,[1]}\right)^2 + \left[x_2^{\,[1]}\right]^2}
\,, \nonumber \\
y_2^{\,[1]}(x_1^{\,[1]},x_2^{\,[1]}) & = & x_2^{\,[1]} - 
\frac{x_2^{\,[1]}}{\left[x_1^{\,[1]}\right]^2+\left[x_2^{\,[1]}\right]^2} - \nonumber \\
& & \quad
- \,q\,\frac{x_2^{\,[1]}}{\left(x_1^{\,[1]} -d^{\,[1]}\right)^2 + \left[x_2^{\,[1]}\right]^2}\,.
\end{eqnarray}
Therefore, for $q \ll 1$, it seems straightforward to characterize the binary lens
as a point-lens at the position of the primary object and treat the secondary
object as perturbation (Gould \& Loeb~\cite{GL}; Gaudi \& Gould~\cite{GG};
Griest \& Safizadeh~\cite{GS}).

For distances from both objects much larger than their separation, however, it
is more appropriate to introduce coordinates $(X_1,X_2)$ and $(Y_1,Y_2)$ 
centered on the center of mass of the lens objects, so that
object 1 
is located at $(d_1,0)$, and object 2 is located at $(-d_2,0)$, where $d_1+d_2 = d$
and
\begin{equation}
d_1 = \frac{q}{1+q}\,d\,,\quad d_2 = \frac{1}{1+q}\,d\,. \label{eq:defd1d2}
\end{equation}
The lens equation in these coordinates reads
\begin{eqnarray}
Y_1 (X_1,X_2) & = & X_1 - \frac{1}{1+q}\,\frac{X_1-\frac{q}{1+q}\,d}{(X_1-\frac{q}{1+q}\,d)^2 + X_2^2} - \nonumber \\
& & \quad - \,\frac{q}{1+q}\,\frac{X_1+\frac{1}{1+q}\,d}
{(X_1+\frac{1}{1+q}\,d)^2+X_2^2}\,, \nonumber \\ 
Y_2 (X_1,X_2) & = & X_2 - \frac{1}{1+q}\,\frac{X_2}{(X_1-\frac{q}{1+q}\,d)^2 + X_2^2} -
\nonumber \\ & & \quad -\,\frac{q}{1+q}\,\frac{X_2}
{(X_1+\frac{1}{1+q}\,d)^2+X_2^2}\,.  
\label{eq:lenseqcom}
\end{eqnarray}
For $X_1 \gg d$, one recovers the lens equation of a point-mass lens at the center
of mass.

\section{The Chang-Refsdal lens with pure shear}
\label{sec:CR}

\begin{figure*}
\resizebox{\hsize}{!}{\includegraphics{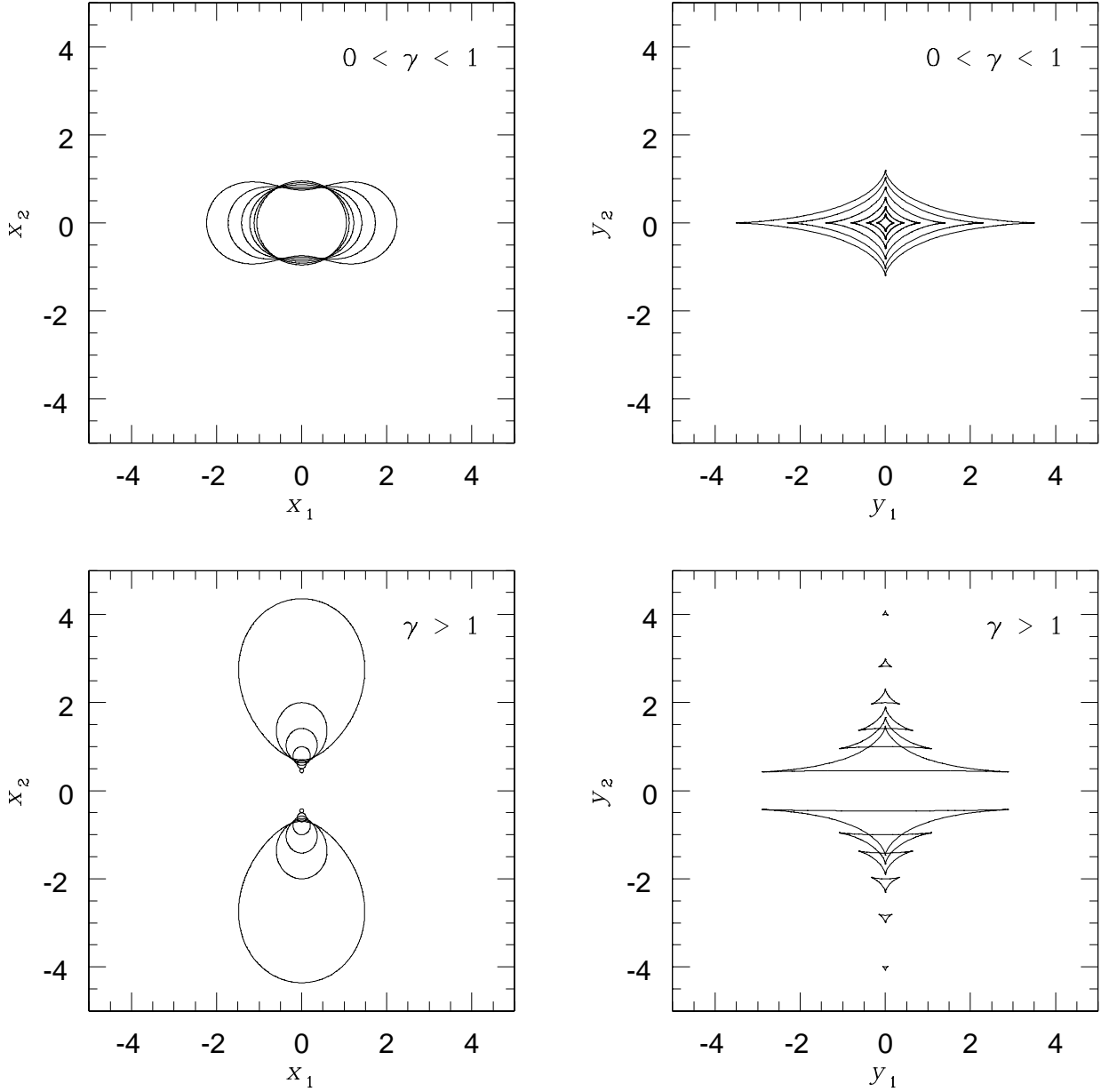}}
\caption{Examples for critical curves (left) and caustics (right)
for a pure-shear Chang-Refsdal lens.
$0 < \gamma < 1$ (top): The
curves shown correspond to $\gamma = 0.1$,  
$0.2$, $1/3$, $0.5$, $2/3$, and $0.8$, where
the critical curves closest to the circle correspond to the smallest $\gamma$ and
larger
caustics correspond to larger values of $\gamma$.
$\gamma > 1$ (bottom): The
curves shown correspond to $\gamma = 1.0526$, $1.25$, 
$1.5$, $2.0$, $3.0$, and $5.0$, where
larger critical curves and larger cautics 
correspond to smaller values of $\gamma$.} 
\label{CRM}
\label{CRm}
\end{figure*}

The lens equation of a Chang-Refsdal lens (Chang \& Refsdal \cite{CRlens}, \cite{CR2}) with pure shear reads
\begin{eqnarray}
y_1(x_1,x_2) & = & (1+\gamma) x_1 - \frac{x_1}{x_1^2+x_2^2}\,, \nonumber \\
y_2(x_1,x_2) & = & (1-\gamma) x_2 - \frac{x_2}{x_1^2+x_2^2}\,, 
\end{eqnarray}
where $|\gamma| \neq 1$, that is the lens equation of a point-lens plus
shear terms.
Since changing the sign of $\gamma$ is equivalent to an interchange
of the coordinate axes, we restrict ourselves here to the case $\gamma > 0$.\footnote{
As we will see in the later sections, binary lens components located
on the $x_2$-axis yield $\gamma > 0$.}
The Jacobian determinant of this mapping
vanishes for
\begin{eqnarray}
& & \left[(1+\gamma)(x_1^2+x_2^2)^2 + x_1^2 - x_2^2\right]\,\cdot \nonumber \\
& & \quad \cdot
\left[(1-\gamma)(x_1^2+x_2^2)^2 - x_1^2 + x_2^2\right] - 4 x_1^2 x_2^2 = 0\,,
\label{CRjac}
\end{eqnarray}
yielding the condition for critical curves.

Solutions on the axes can be found easily. For $x_1 =0$, Eq.~(\ref{CRjac}) reduces to
\begin{equation}
\left[(1+\gamma)x_2^2 - 1\right]\,
\left[(1-\gamma)x_2^2 + 1\right] = 0 \,,
\end{equation}
which directly yields the solutions $(0,\pm x_2^{(\mathrm{a})})$ where
\begin{equation}
x_2^{(\mathrm{a})} = \frac{1}{\sqrt{1+\gamma}}
\end{equation}
for all positive $\gamma \neq 1$, and $(0,\pm x_2^{(\mathrm{b})})$ where
\begin{equation}
x_2^{(\mathrm{b})} = \frac{1}{\sqrt{\gamma-1}}
\end{equation}
for $\gamma > 1$. 

Since $x_1 = 0$ maps to $y_1 =0$, one immediately obtains the intersections of the caustics with the $y_2$-axis
which are at $(0,\pm y_2^{(\mathrm{r})})$, where 
$y_2^{(\mathrm{r})} = y_2(0,-x_2^{(\mathrm{r})})$ ($\mathrm{r} = \mathrm{a},\mathrm{b}$) and
\begin{equation}
y_2^{(\mathrm{a})} = \frac{2 \gamma}{\sqrt{1+\gamma}}\,, \quad
y_2^{(\mathrm{b})} = 2\,\sqrt{\gamma-1}\,.
\end{equation}

For $x_2 = 0$, one obtains
\begin{equation}
\left[(1-\gamma)x_1^2 - 1\right]\,
\left[(1+\gamma)x_1^2 + 1\right] = 0 \,,
\end{equation}
which, for positive $\gamma$, yields the solutions $(\pm x_1^{(\mathrm{a})},0)$ where
\begin{equation}
x_1^{(\mathrm{a})} = \frac{1}{\sqrt{1-\gamma}}
\end{equation}
for $0 < \gamma < 1$.
The critical points $(\pm x_1^{(\mathrm{a})},0)$ are mapped onto the intersections of the caustic with the
$y_1$-axis at $(\pm y_1^{(\mathrm{a})},0)$, where
\begin{equation}
y_1^{(\mathrm{a})} = \frac{2 \gamma}{\sqrt{1-\gamma}}\,.
\end{equation}

Let us now discuss the two cases $\gamma < 1$ and $\gamma > 1$.
For $\gamma < 1$,
there is one critical curve which intersects the $x_1$-axis at
$(\pm x_1^{(\mathrm{a})},0)$ and the $x_2$-axis at $(0,\pm x_2^{(\mathrm{a})})$. For
$\gamma \to 0$,
\begin{eqnarray}
x_1^{(\mathrm{a})} & \simeq & 1 + \frac{\gamma}{2}\,, \nonumber \\
x_2^{(\mathrm{a})} & \simeq & 1 - \frac{\gamma}{2}\,, 
\end{eqnarray}
so that the critical curve tends to a circle with radius 1 for $\gamma \to 0$.
For $\gamma \to 1$, $x_2^{(\mathrm{a})} \to 1/\sqrt{2}$, while
$x_1^{(\mathrm{a})} \to \infty$, so that the size of the critical curve
grows to infinity (like $(1-\gamma)^{-1/2}$).
The critical curve is mapped to a
diamond shaped caustic which intersects the $y_1$-axis at 
$(\pm y_1^{(\mathrm{a})},0)$ and the $y_2$-axis at $(0, \pm y_2^{(\mathrm{a})})$. These intersections are
the positions of the 4 cusps of the caustic. The extent of the caustic along the $y_1$-axis ($s_1$) and
along the $y_2$-axis ($s_2$) is given by
\begin{eqnarray}
s_1 & = & 2 y_1^{(\mathrm{a})} = 
\frac{4\gamma}{\sqrt{1-\gamma}} \simeq 4 \gamma + 2 \gamma^2
\quad (\gamma \ll 1)\,, \\ 
s_2 & = & 2 y_2^{(\mathrm{a})} = 
\frac{4\gamma}{\sqrt{1+\gamma}} \simeq 4 \gamma - 2 \gamma^2
\quad (\gamma \ll 1)\,.
\end{eqnarray} 
Note that $s_1 > s_2$, and $s_i \sim 4\gamma$, $s_1 -s_2 \sim 4 \gamma^2$ for $\gamma \ll 1$.
Further note that the size of the caustic increases with $\gamma$ and for $\gamma \to 1$, 
$s_1 \to \infty$ and $s_2 \to 2 \sqrt{2}$.
Examples for the shape of the critical curves and caustics 
for $0 < \gamma < 1$ are shown in Fig.~\ref{CRM}
 (see also Chang \& Refsdal~\cite{CR2}).

For $\gamma > 1$, there are no intersections of the critical curves with the
$x_1$-axis, but 4 intersections with the $x_2$-axis, namely at
$(0,\pm x_2^{(\mathrm{a})})$ and $(0,\pm x_2^{(\mathrm{b})})$. There are two critical curves,
where one of them intersects the positive and the other one the negative
$x_2$-axis. For $\gamma \to \infty$, one obtains
\begin{eqnarray}
x_2^{(\mathrm{a})} = \frac{1}{\sqrt{\gamma}}\,\left(1-\frac{1}{2 \gamma}\right) \,, \nonumber \\
x_2^{(\mathrm{b})} = \frac{1}{\sqrt{\gamma}}\,\left(1+\frac{1}{2 \gamma}\right) \,, 
\end{eqnarray}
so that $x_2^{(\mathrm{a})}$ and $x_2^{(\mathrm{b})}$ tend to zero like $1/\sqrt{\gamma}$ and 
the size of the critical curves (between $x_2^{(\mathrm{b})}$ and $x_2^{(\mathrm{a})}$) tends to
zero like $\gamma^{-3/2}$. For $\gamma \to 1$, $x_2^{(\mathrm{a})} \to 1/\sqrt{2}$, while
$x_2^{(\mathrm{b})} \to \infty$, so that, as for $0 < \gamma < 1$, the size of the 
critical curves grows to infinity (like $(\gamma -1)^{-1/2}$).
Like the critical curves, the caustics 
do not intersect the $y_1$-axis, but there are four intersections with
the $y_2$-axis, namely at $(0, \pm y_2^{(\mathrm{a})})$ and at 
$(0, \pm y_2^{(\mathrm{b})})$. There are two caustics shaped like a triangle, each with 3 cusps, where
one cusp is at   
$(0, \pm y_2^{(\mathrm{a})})$ pointing towards larger $|y_2|$. The inner distance between the
triangular shaped caustics is given by
\begin{equation}
\Delta_\mathrm{i} = 2 y_2^{(\mathrm{b})} = 
4\,\sqrt{\gamma-1} \simeq 4 \sqrt{\gamma}\left(1-\frac{1}{2\gamma} +
\frac{3}{8 \gamma^2}\right) \,, 
\end{equation}
and the outer distance (between the cusps on the $y_2$-axis) is
\begin{equation} 
\Delta_\mathrm{o} = 2 y_2^{(\mathrm{a})} =
\frac{4 \gamma}{\sqrt{1+\gamma}} \simeq
4 \sqrt{\gamma}\left(1-\frac{1}{2\gamma}-\frac{1}{8 \gamma^2}\right) \,,
\end{equation} 
where the expansions are for $\gamma \gg 1$.
The 'size' of the caustics in $y_2$-direction is given by
\begin{equation}
s_2 = \frac{1}{2}\,(\Delta_\mathrm{o} - \Delta_\mathrm{i}) = 
\frac{2 \gamma}{\sqrt{1+\gamma}} - 2 \sqrt{\gamma -1} \simeq \gamma^{-3/2}\,.
\end{equation}
Note that the size of the caustics decreases with larger $\gamma$. For $\gamma \gg 1$,
the size of the caustics is $\gamma^{-3/2}$ and the distance between the two caustics is $4 \sqrt{\gamma}$.
For $\gamma \to 1$, the extent of the caustics in $y_1$-direction goes to infinity, while 
$\Delta_\mathrm{i} \to 0$ and $\Delta_\mathrm{o} \to 2 \sqrt{2}$, which meets the behaviour of the case
$\gamma < 1$ for $\gamma \to 1$. 
Examples for the shape of the critical curves and caustics  for $\gamma > 1$
are shown in Fig.~\ref{CRm}
(see also Chang \& Refsdal~\cite{CR2}).

\section{The quadrupole lens}
\label{sec:quad}

\begin{figure*}
\resizebox{12cm}{!}{\includegraphics{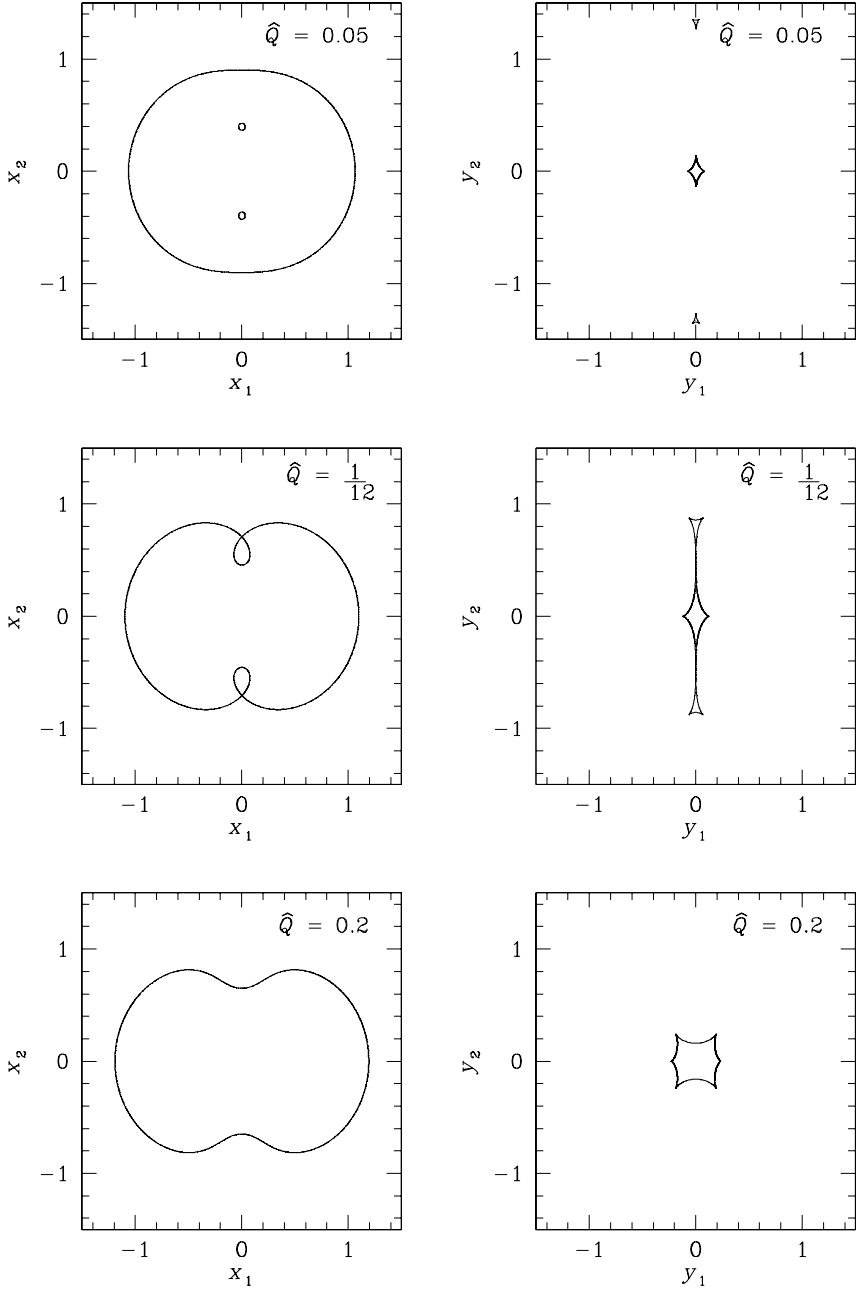}}
\hfill
\parbox[b]{55mm}{
\caption{The shapes of 
critical curves (left) and caustics (right) of a quadrupole lens for
$\hat Q < \frac{1}{12}$, $\hat Q = \frac{1}{12}$, and $\hat Q > \frac{1}{12}$.
There are some additional transitions for the caustics for $\hat Q \geq 0.25$ which
are not shown here.}
\label{fig:quad}}
\end{figure*}

For a distance from the lens masses that is much larger than the distance
between them, the deflection potential can be approximated by means of the
multipole expansion (e.g.\ Landau \& Lifshitz~\cite{Landau}; Jackson~\cite{Jackson}).

By expanding the term
$f(\vec x - \vec x^{(r)}) = \ln \left|\vec x - \vec x^{(r)}\right|$ 
in the deflection potential $\psi$ (see Eq.~(\ref{eq:psi})) as
\begin{eqnarray}
f(\vec x - \vec x^{(r)}) & \simeq & f(\vec x) - (\vec \nabla f)(\vec x)\cdot
\vec x^{(r)} + \nonumber \\ & & +\, \frac{1}{2}
\sum_{i=1}^{2} \sum_{j=1}^{2} \frac{\partial^2 f}{\partial x_i \partial x_j}
(\vec x)\,x_i^{(r)}\,x_j^{(r)}\,,
\end{eqnarray}
one obtains for $\psi$
\begin{eqnarray}
\psi(\vec x) & \simeq & 
\ln |\vec x| - \frac{1}{|\vec x|^2}
\sum_{r=1}^{N} m_r\, \vec x \cdot\vec x^{(r)}
+ \nonumber \\ 
& & +\,\frac{1}{2}\,\sum_{r=1}^{N}\sum_{i=1}^{2}\sum_{j=1}^{2}
\frac{|\vec x|^2\,\delta_{ij} - 2 x_i x_j}{|\vec x|^4}
\,m_r\, x_i^{(r)} x_j^{(r)}\,.
\end{eqnarray}
With
\begin{eqnarray}
 & & \sum_{i=1}^{2}\sum_{j=1}^{2} |\vec x|^2\,\delta_{ij}\, x_i^{(r)} x_j^{(r)}
 = |\vec x|^2\,|\vec x^{(r)}|^2 \nonumber \\
& & \quad = \sum_{i=1}^{2}\sum_{j=1}^{2} |\vec x^{(r)}|^2\,\delta_{ij}\, x_i x_j
\end{eqnarray}
the deflection potential finally reads
\begin{equation}
\psi(\vec x) = \ln |\vec x| - \frac{\vec x \cdot \vec p}{|\vec x|^2}
+ \frac{1}{2}\,\frac{1}{|\vec x|^4}\, \sum_{i=1}^{2} \sum_{j=1}^{2} x_i x_j Q_{ij}\,,
\label{eq:quadpot}
\end{equation}
where $\vec p$ is the {\em dipole moment}
\begin{equation}
\vec p = \sum_{r=1}^{N} m_r\,\vec x^{(r)}\,,
\end{equation}
and $Q_{ij}$ denotes the components of the {\em quadrupole moment} $Q$
\begin{equation}
Q_{ij} = \sum_{r=1}^{N} m_r\,(|\vec x^{(r)}|^2\,\delta_{ij} - 2 
x_i^{(r)} x_j^{(r)})\,,
\end{equation}
where $Q$ is symmetric and traceless, i.e. $Q_{11} = - Q_{22}$ and
$Q_{12} = Q_{21}$. From the deflection potential $\psi$ given in Eq.~(\ref{eq:quadpot}) and
from Eqs.~(\ref{eq:lenseqgen}) and~(\ref{eq:gradpot}), one obtains the lens equation
\begin{eqnarray}
\vec y & = & \vec x - \frac{\vec x}{|\vec x|^2}
+ \frac{|\vec x|^2 \vec p - 2 (\vec x\cdot \vec p) \vec x}{|\vec x|^4}
+ \nonumber \\
 & & \quad +\,\frac{2 (\vec x^{\rm T} Q \vec x) \vec x - |\vec x|^2 Q \vec x}{|\vec x|^6}
\,.
\label{eq:lenseqquad0}
\end{eqnarray}
By choosing the center of mass being the coordinate origin, the 
dipole moment vanishes. For a binary lens with object 2 at $(-d_2,0)$ and
object 1 at $(d_1,0)$,
Eqs.~(\ref{eq:defq}) 
and~(\ref{eq:defd1d2})
yield the quadrupole moment  
\begin{eqnarray}
\hat Q & = & Q_{22} = - Q_{11} = m_1 d_1^2 + m_2 d_2^2 = \frac{q\,d^2}{(1+q)^2} > 0\,,
\nonumber \\
\quad Q_{12} & = & Q_{21} = 0\,.
\label{eq:quadmoment1}
\end{eqnarray}
For $q \ll 1$, one obtains
\begin{equation}
\hat Q = \frac{q}{(1+q)^3}\,\left[d^{\,[1]}\right]^2
\simeq q \left[d^{\,[1]}\right]^2
\simeq q^2 \left[d^{\,[2]}\right]^2\,,
\label{eq:quadmoment}
\end{equation}
using the angular separations $d^{\,[1]}$ and $d^{\,[2]}$ in units of 
$\theta_{\rm E}^{\,[1]}$ and $\theta_{\rm E}^{\,[2]}$ as defined in Eq.~(\ref{eq:d1d2def}).
Since $Q$ is symmetric and traceless, the eigenvalues of $Q$ differ only in sign, and $\hat Q$ is
their absolute value.
From Eqs.~(\ref{eq:lenseqquad0}) and~(\ref{eq:quadmoment1}), the
lens equation for the quadrupole limit of the binary lens reads
\begin{eqnarray}
y_1(x_1,x_2) & = & x_1 - \frac{x_1}{x_1^2+x_2^2}
+ \hat Q\, \frac{3 x_1 x_2^2 - x_1^3}{(x_1^2+x_2^2)^3}\,, \nonumber \\
y_2(x_1,x_2) & = & x_2 - \frac{x_2}{x_1^2+x_2^2}
+ \hat Q\, \frac{x_2^3 - 3 x_1^2 x_2}{(x_1^2+x_2^2)^3} \,,
\end{eqnarray}
and the derivatives of the lens mapping are
\begin{eqnarray}
\frac{\partial y_1}{\partial x_1}(x_1,x_2) & = & 1 
+ \frac{x_1^2-x_2^2}{(x_1^2+x_2^2)^2} + \nonumber \\
& & \quad +\,\frac{3 \hat Q}
{(x_1^2+x_2^2)^4}\left(x_1^4 + x_2^4 - 18 x_1^2 x_2^2\right)\,, \nonumber \\
\frac{\partial y_2}{\partial x_2}(x_1,x_2) & = & 1 
- \frac{x_1^2-x_2^2}{(x_1^2+x_2^2)^2} - \nonumber \\
& & \quad -\,\frac{3 \hat Q}
{(x_1^2+x_2^2)^4}\left(x_1^4 + x_2^4 - 18 x_1^2 x_2^2\right)\,, \nonumber \\
\frac{\partial y_1}{\partial x_2}(x_1,x_2) & = &  
  \frac{\partial y_2}{\partial x_1}(x_1,x_2)  =  
\frac{2 x_1 x_2}{(x_1^2+x_2^2)^2} - \nonumber \\
& & \quad -\,\frac{12 \hat Q}
{(x_1^2+x_2^2)^4}\left
(x_1^3 x_2 - x_1 x_2^3\right)\,. 
\end{eqnarray}

From the derivatives, the 
intersections of the critical curves with the axes can be obtained as
follows. For $x_1 = 0$, one obtains the condition
\begin{equation}
\left[1-\frac{1}{x_2^2}+\frac{3 \hat Q}{x_2^4}\right]
\left[1+\frac{1}{x_2^2}-\frac{3 \hat Q}{x_2^4}\right] = 0\,,
\label{eq:x1eq0}
\end{equation}
which is fulfilled if one of the factors becomes zero.
Satisfying this condition with the first factor yields
\begin{equation}
x_2^2 = \frac{1}{2}\left(1 \pm \sqrt{1-12 \hat Q}\right)\,.
\end{equation}
For $\hat Q > \frac{1}{12}$, there is no solution; for $\hat Q = \frac{1}{12}$,
there is the solution $x_2 = \pm\frac{1}{2}\,\sqrt{2}$, while for
$\hat Q < \frac{1}{12}$, there are the two solutions
\begin{equation}
{x_2^{(\mathrm{a})}}^2 = \frac{1}{2}\left(1+\sqrt{1-12 \hat Q}\right) \simeq 1 - 3 \hat Q - 9 \hat
Q ^2 \; (\hat Q \ll 1)\,,
\end{equation}
and
\begin{equation}
{x_2^{(\mathrm{b})}}^2 = \frac{1}{2}\left(1-\sqrt{1-12 \hat Q}\right) \simeq 3 \hat Q + 9 \hat
Q ^2 \quad (\hat Q \ll 1)\,.
\end{equation}

Satisfying the condition of Eq.~(\ref{eq:x1eq0}) with the second factor gives 
\begin{equation}
x_2^2 = - \frac{1}{2}\left(1 \pm \sqrt{1 + 12 \hat Q}\right)\,,
\end{equation}
where only
\begin{equation}
{x_2^{(\mathrm{c})}}^2 = \frac{1}{2}\left(\sqrt{1+12 \hat Q}-1\right) \simeq 3 \hat Q - 9 \hat
Q ^2 \quad (\hat Q \ll 1)\,,
\end{equation}
yields a solution ($\hat Q > 0$).

Along the other axis ($x_2 = 0$), one obtains the condition
\begin{equation}
\left[1+\frac{1}{x_2^2}+\frac{3 \hat Q}{x_2^4}\right]
\left[1-\frac{1}{x_2^2}-\frac{3 \hat Q}{x_2^4}\right] = 0\,.
\end{equation}
Satisfying this condition with the first factor gives
\begin{equation}
x_1^2 = \frac{1}{2}\left(\sqrt{1-12 \hat Q}-1\right)\,,
\end{equation}
which does not yield solutions for $\hat Q > 0$, while the
second factor gives
\begin{equation}
x_1^2 = \frac{1}{2}\left(1 \pm \sqrt{1+12 \hat Q}\right)\,,
\end{equation}
yielding the solution
\begin{equation}
{x_1^{(\mathrm{a})}}^2 = \frac{1}{2}\left(1+\sqrt{1+12 \hat Q}\right) \simeq 1 + 3 \hat Q - 9 \hat
Q ^2 \; (\hat Q \ll 1)\,.
\end{equation}

The points $(\pm x_1^{(\mathrm{a})},0)$ and $(0,\pm x_2^{(\mathrm{a})})$ are the intersections of the axes  
with a critical curve near the Einstein ring. This critical curve is mapped to a diamond-shaped
caustic with 4 cusps on the coordinate axes at $(\pm y_1^{(\mathrm{a})},0)$ and
$(0,\pm y_2^{(\mathrm{a})})$, where
\begin{eqnarray}
y_1^{(\mathrm{a})} & = & y_1(x_1^{(\mathrm{a})},0) \simeq 2 \hat Q + 9 \hat Q^2\quad (\hat Q \ll 1)\,, \nonumber \\
y_2^{(\mathrm{a})} & = & y_2(0,-x_2^{(\mathrm{a})}) \simeq 2 \hat Q - 9 \hat Q^2\quad (\hat Q \ll 1)\,. 
\end{eqnarray}
This shows that for $\hat Q \ll 1$, the size of the caustic
is $4 \hat Q$, where the extent along the $y_2$-axis exceeds the extent along the
$y_1$-axis by $18 \hat Q^2$.

The points $(0,\pm x_2^{(\mathrm{b})})$ and $(0,\pm x_2^{(\mathrm{c})})$ describe 
intersections with the $x_2$-axis of 
small closed critical curves in that region, corresponding to triangular-shaped caustics
on the $y_2$-axis intersecting the axis at
\begin{eqnarray}
y_2^{(\mathrm{b})}  & = & y_2(0,-x_2^{(\mathrm{b})}) \nonumber \\
& \simeq &
\frac{2}{9}\sqrt{3}\hat Q^{-1/2} - \sqrt{3}\,\hat Q^{1/2} - \frac{9}{4}\sqrt{3} \hat Q^{3/2}
\quad (\hat Q \ll 1)\,, \nonumber \\
y_2^{(\mathrm{c})}  & = & y_2(0,-x_2^{(\mathrm{c})}) \nonumber \\
& \simeq &
\frac{2}{9}\sqrt{3}\hat Q^{-1/2} - \sqrt{3}\,\hat Q^{1/2} + \frac{3}{4}\sqrt{3} \hat Q^{3/2}
\quad (\hat Q \ll 1)\,.
\end{eqnarray}
Thus the size of these caustics along the $y_2$-axis is $3\sqrt{3} \hat Q^{3/2}$
for small $\hat Q$, and tends to zero for
small $\hat Q$, while the caustics themselves move to infinity 
as $\hat Q^{-1/2}$.

Fig.~\ref{fig:quad} illustrates the topologies of the critical curves and the caustics. The critical
curves have two
different topologies: a single curve for $\hat Q > \frac{1}{12}$ and an outer
curve with two smaller inner curves for $\hat Q < \frac{1}{12}$. The outer curve for
$\hat Q < \frac{1}{12}$ and the only curve for $\hat Q > \frac{1}{12}$ intersect the axes at
$(\pm x_1^{(\mathrm{a})},0)$ and
$(0,\pm x_2^{(\mathrm{a})})$, while the inner curves intersect the $x_2$-axis at
$(0,\pm x_2^{(\mathrm{b})})$ and
$(0,\pm x_2^{(\mathrm{c})})$.
The corresponding caustics for 
$\hat Q < \frac{1}{12}$ are a diamond-shaped caustic at the origin and two triangular-shaped caustics
along the $y_2$-axis symmetrically placed with respect to the $y_1$-axis. 
At $\hat Q = \frac{1}{12}$, these caustics merge
to form a six-cusp caustic as the two inner critical curves touch the
outer curve at $(0,\pm \sqrt{2}/2)$. Though the only change in the shape of the critical curves for $\hat Q >
\frac{1}{12}$ is that the dumb-bell shape transforms into an oval shape, there are some transitions in
the shape of the caustics. For $\hat Q = 0.25$, the caustic touches itself at the origin and becomes
self-intersecting for $\hat Q > 0.25$. At the same time, the left and right caustics break up into 3
caustics each, so that the six-cusp caustic becomes a self-intersecting 10-cusp caustic. After a few other
transitions,  a two-cusp caustic arises, with the two cusps on the $x_1$-axis and pointing inwards.

\section{The different topologies of the critical curves and caustics of the binary lens}
\label{sec:bintop}

\begin{figure}
\resizebox{\hsize}{!}{\includegraphics{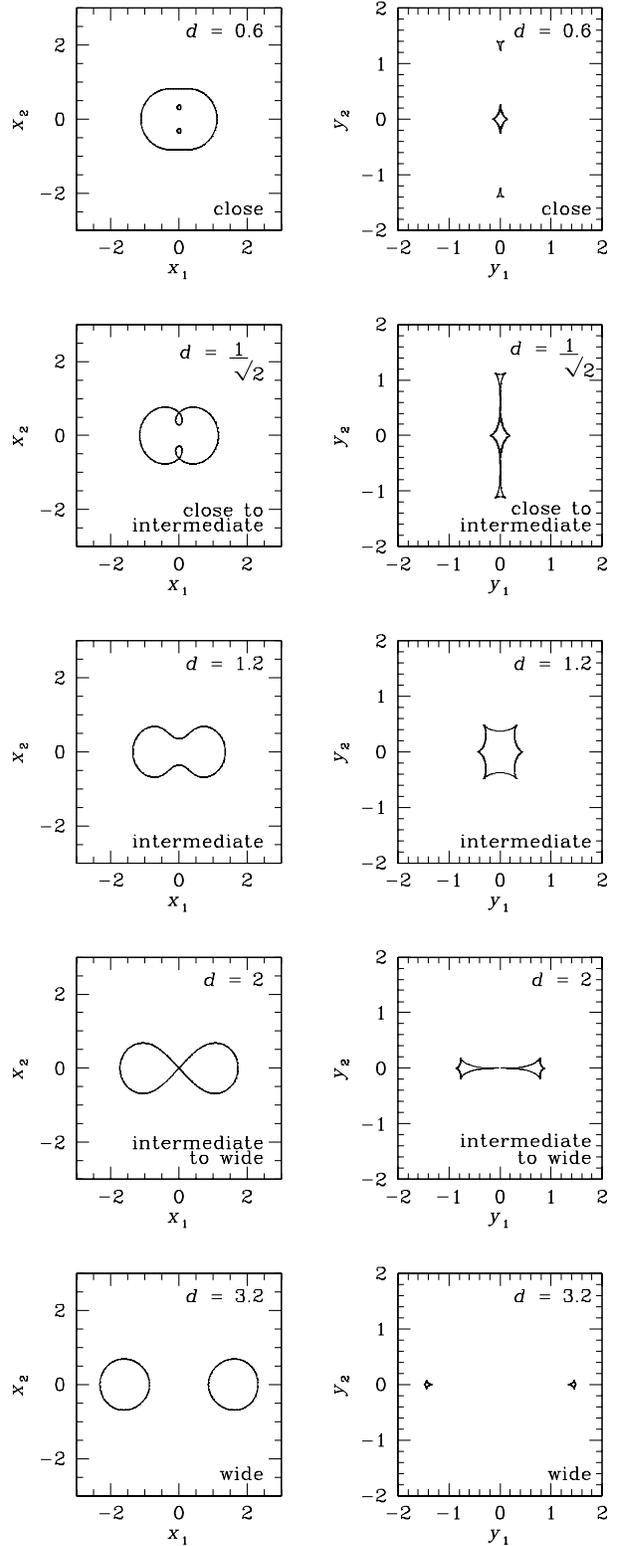}}
\caption{The 3 different topologies of the critical curves (left) and
caustics (right) of a binary lens with $q=1$: close, intermediate, and wide.
The transitions occur at $d = 1/\sqrt{2}$ and $d = 2$.}
\label{ccbin}
\label{causticbin}
\end{figure}

For a mass ratio $q=1$, Schneider \& Wei{\ss} (\cite{SchneiWei}) have shown
that 3 different 
topologies of the critical curves and the associated caustics exist, which are named
'close binaries', 'intermediate binaries', and 'wide binaries' in this paper.
Close binaries correspond to separations
$d < 1/\sqrt{2}$ and have one large critical curve with 2 smaller critical
curves inside it, positioned 
on the $x_2$-axis symmetrically about the $x_1$-axis. The large critical curve
is mapped to a central diamond-shaped
caustic with 4 cusps, while the 2 small critical curves
are mapped to two triangular-shaped caustics with 3 cusps each, located along the
$y_2$-axis and symmetrically to the $y_1$-axis.
For $d \to 0$, the small
critical curves become smaller and move towards the $x_1$-axis, while the larger critical curve tends
to a circle of unit radius. In the same limit, the triangular-shaped caustics
become smaller and move to infinity, while the diamond-shaped caustic degenerates into a point at the origin.
At $d = 1/\sqrt{2}$, the 2 inner critical curves touch the large critical curve to form a
single dumb-bell shaped curve, and
the diamond-shaped caustic and the triangular-shaped caustics touch with two of their
cusps on each side, creating one caustic with 6 cusps for 
$1/\sqrt{2} < d < 2$, which marks the regime of intermediate binaries.
At the boundary between intermediate and wide binaries at $d = 2$, 
the critical curve touches itself at the origin and separates into two curves
for $d > 2$, while the 
six-cusp caustic separates at the origin into two diamond-shaped caustics.
The wide binary regime is entered for
$d > 2$, where the critical curves tend to 
circles with radius $1/\sqrt{2}$ around each
lens object, while the caustics
move towards the positions of the two lens objects, 
and degenerate into point caustics
for $d \to \infty$. The shape of the critical curves and caustics for the different topologies is 
illustrated in Fig.~\ref{ccbin}.

\begin{figure}[t]
\resizebox{\hsize}{!}{\includegraphics{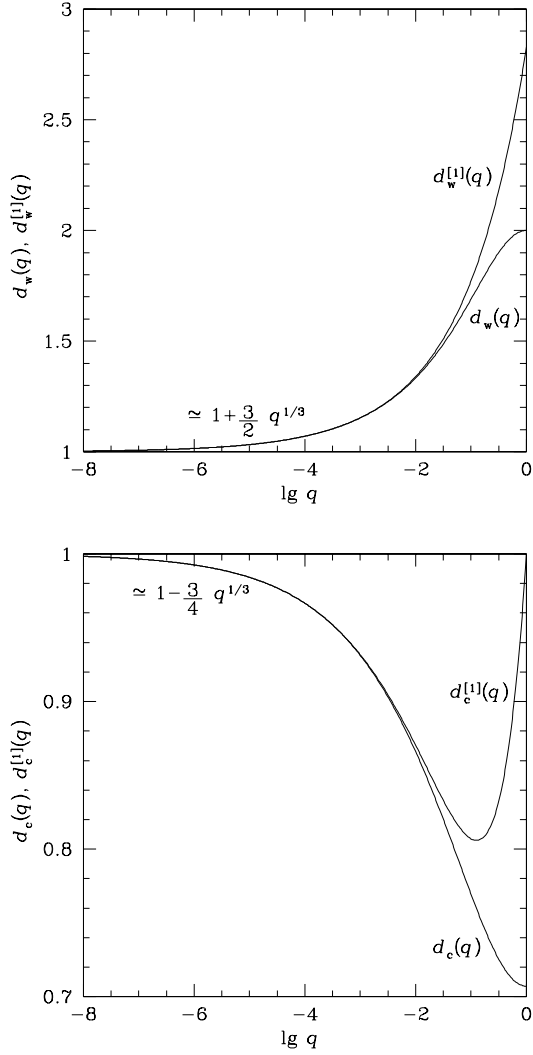}}
\caption{The bifurcation values between the 3 different topologies of the caustics (and critical curves) 
as a function of the mass ratio $q$.
The top panel shows the bifurcation values between
intermediate and wide binaries ($d_\mathrm{w}$,
$d_\mathrm{w}^{\,[1]}$). For $q = 1$,
$d_\mathrm{w}(q=1) = 2$, and 
$d_\mathrm{w}^{\,[1]}(q=1) = 2\,\sqrt{2}$.
The bottom panel shows the 
bifurcation values between intermediate and close binaries ($d_\mathrm{c}$, 
$d_\mathrm{c}^{\,[1]}$).  For $q = 1$, $d_\mathrm{c}(q = 1) = \sqrt{2}/2$ and
$d_\mathrm{c}^{\,[1]}(q=1) = 0$, and the minimum of $d_\mathrm{c}^{\,[1]}(q)$ is at
$q = 1/8$ and $d_\mathrm{c}^{\,[1]} = (3/4)^{3/4} \approx 0.806$.
$d_\mathrm{c}$ and $d_\mathrm{w}$ measure the angular separation
between the primary object and the secondary object in units of $\theta_\mathrm{E}$,
while $d_\mathrm{c,w}^{\,[1]} = \sqrt{1+q}\,d_\mathrm{c,w}$ measure the
angular separation in units of $\theta_\mathrm{E}^{\,[1]}$.}
\label{dclose}
\label{dwide}
\end{figure}

Erdl \& Schneider (\cite{Erdl}) have shown that the topologies of the 
critical curves and caustics remain the same for any arbitrary mass
ratio. However, the critical curves and 
caustics are stretched and shifted in position, and have different size. 
Furthermore, the separations at which the transitions between
close, intermediate, and wide binaries occur depend on the mass ratio. 
For the general case, the bifurcation values $d_\mathrm{c}$ between close and intermediate 
binaries  
and $d_\mathrm{w}$ between intermediate and wide binaries are given by (Erdl \& Schneider~\cite{Erdl})
\begin{equation}
m_1\,m_2 = \frac{(1-d_\mathrm{c}^4)^3}{27 d_\mathrm{c}^8}\,,
\end{equation}
and
\begin{equation}
d_\mathrm{w}  = \left(m_1^{1/3}+m_2^{1/3}\right)^{3/2}\,.
\end{equation}

By means of the mass ratio $q$ and using separations in units of $\theta_\mathrm{E}^{\,[1]}$, the bifurcation
values can be written as
\begin{equation}
d_\mathrm{c}^{\,[1]}(q)  = \left\{\frac{\left[(1+q)^2-{d_\mathrm{c}^{\,[1]}}^4\right]^3}{27 q}\right\}^{1/8}
\,,\footnotemark
\end{equation}
\footnotetext{Note that this is a fix-point equation for
$d_\mathrm{c}^{\,[1]}$, which can easily be iterated.}
and 
\begin{equation}
d_\mathrm{w}^{\,[1]}(q) = \left(1+q^{1/3}\right)^{3/2}\,.
\end{equation}

Fig.~\ref{dclose} shows $d_\mathrm{c}^{\,[1]}(q)$, $d_\mathrm{c}(q)$,
$d_\mathrm{w}^{\,[1]}(q)$, and $d_\mathrm{w}(q)$.
One sees that $d_\mathrm{w}^{\,[1]} > d_\mathrm{w} > 1$ for all $0 < q \leq 1$,
so that there are no wide binaries for $d^{\,[1]} < 1$ or
$d < 1$, and that 
$d_\mathrm{w}^{\,[1]},d_\mathrm{w} \to 1$ for $q \to 0$, so that
all binaries with $d^{\,[1]} > 1$ or
$d > 1$ are wide binaries in the limit $q \to 0$.
$d_\mathrm{w}^{\,[1]}$ and $d_\mathrm{w}$ increase monotonically to
a maximum at $q = 1$, where $d_\mathrm{w}(1) = 2$ and $d_\mathrm{w}^{\,[1]}(1) = 2\,\sqrt{2}$, i.e.\ all binaries with $d^{\,[1]} > 2$ or $d > 2\,\sqrt{2}$ are wide binaries.
For the transition between close and intermediate binaries,
$d_\mathrm{c} < d_\mathrm{c}^{\,[1]} < 1$ for all $0 < q \leq 1$, so that
close binaries occur only for $d^{\,[1]} < 1$ or $d < 1$, and  
$d_\mathrm{c}^{\,[1]},d_\mathrm{c} \to 1$ for $q \to 0$, so that
in this limit, all binaries with $d^{\,[1]} < 1$ or $d < 1$ are close binaries. 
While $d_\mathrm{c}$ decreases monotonically to
$d_\mathrm{c}(1) = \sqrt{2}/2$, $d_\mathrm{c}^{\,[1]}$ goes through a minimum of 
$(3/4)^{3/4} \approx 0.806$ for $q = 1/8$ and reaches
$d_\mathrm{c}^{\,[1]}(1) = 1$, so that all binaries with $d^{\,[1]} < (3/4)^{3/4}$ or
$d < \sqrt{2}/2$ are close binaries. 

It can be seen that the bifurcation values for equal masses correspond to the lens objects being located
on the angular Einstein rings of the individual objects ($d_\mathrm{c}$) or the angular Einstein rings of the total mass
located centered in the middle of the two lens objects ($d_\mathrm{w}$).

For small $q$, $d_\mathrm{c}^{\,[1]}$ and $d_\mathrm{w}^{\,[1]}$ can be approximated by
\begin{eqnarray}
d_\mathrm{c}^{\,[1]}(q) & \simeq & 1-\frac{3}{4}\, q^{1/3}\,,\label{approxdc} \nonumber\\
d_\mathrm{w}^{\,[1]}(q) & \simeq & 1+\frac{3}{2}\, q^{1/3}\,,\label{approxdw}
\end{eqnarray}
and the same
expressions also hold for $d_\mathrm{c}(q)$ and $d_\mathrm{w}(q)$ in this limit.
For small mass ratios, the region for intermediate binaries is small, and the
size of this region decreases with $q^{1/3}$.

For a given separation $d^{\,[1]}$, the critical mass ratios $q_\mathrm{c}$ and $q_\mathrm{w}$, 
for which the transitions between close and intermediate binaries or intermediate and
wide binaries occur, follow from
\begin{equation}
q_\mathrm{c}(d^{\,[1]}) = \frac{\left[(1+q_\mathrm{c})^2-{d^{\,[1]}}^4\right]^3}{27 {d^{\,[1]}}^8}
\end{equation}
and
\begin{equation}
q_\mathrm{w}(d^{\,[1]}) = \left({d^{\,[1]}}^{2/3} - 1\right)^3\,.
\end{equation}
Note that for $(3/4)^{3/4} < d^{\,[1]} < 1$, there are two values for $q_\mathrm{c}$,
while for $d^{\,[1]} < (3/4)^{3/4}$, no solution for $q_\mathrm{c}$ exists and
all mass ratios give close binaries.
Similarly, for $d^{\,[1]} > \sqrt{2}$, one has
a wide binary for any mass ratio.
The approximative relations for $q \ll 1$, Eq.~(\ref{approxdc}), 
can be inverted to yield
\begin{eqnarray}
q_\mathrm{c}(d^{\,[1]}) & \simeq & \frac{64}{27}\,\left(1-d^{\,[1]}\right)^3\,, \nonumber \\
q_\mathrm{w}(d^{\,[1]}) & \simeq & \frac{8}{27}\,\left(d^{\,[1]}-1\right)^3\,.
\end{eqnarray}

\section{The vicinity of the secondary object}
\label{sec:vicsec}

\begin{figure*}
\resizebox{12cm}{!}{\includegraphics{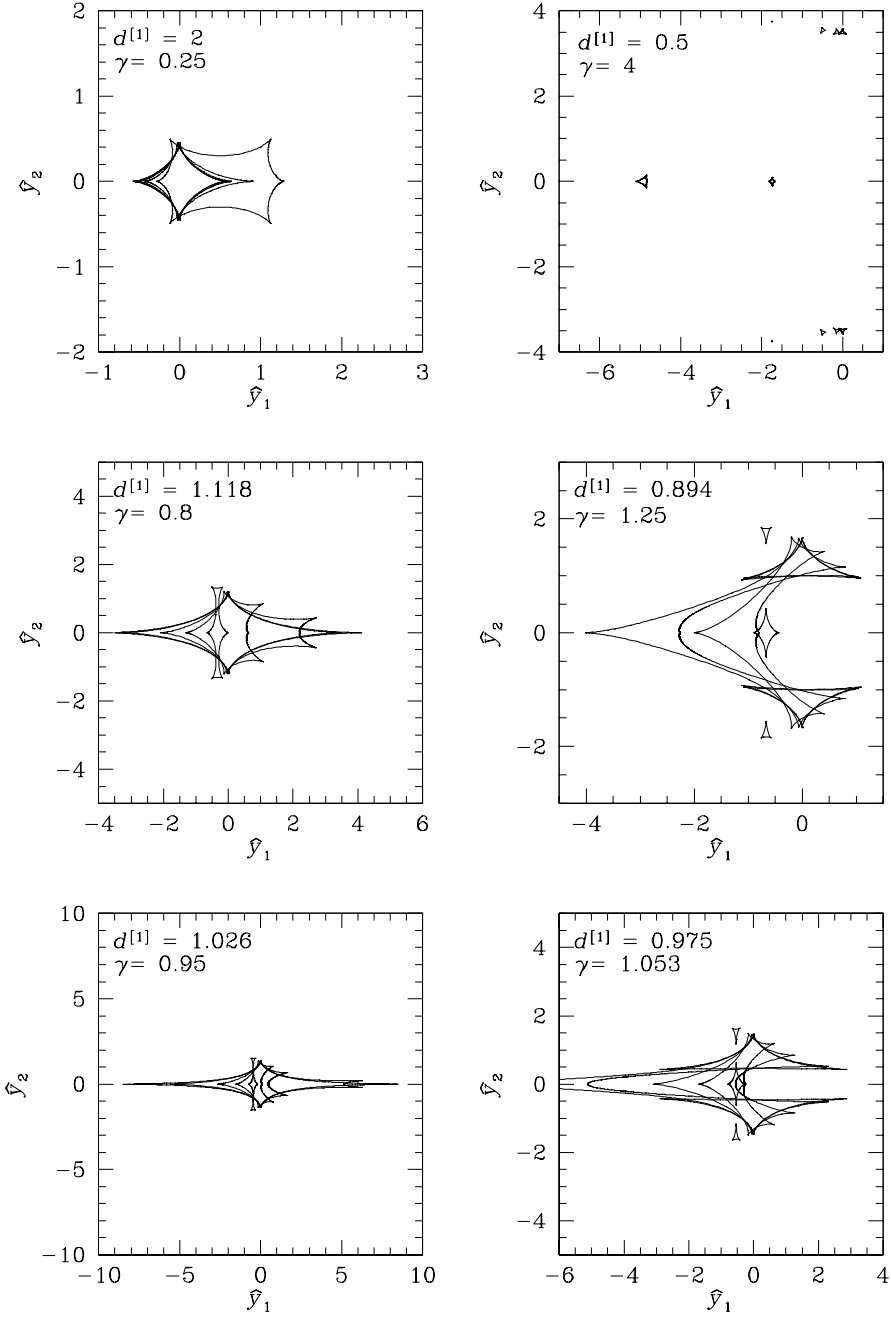}}
\hfill
\parbox[b]{55mm}{
\caption{The transition of the caustic shape from the equal mass case to
extreme mass ratio, or: How the Chang-Refsdal limit is approached for different
separations between primary object and secondary object.
All subfigures show the mass ratios $q = 1$, $0.1$, $0.01$,
$10^{-4}$, and 0 (exact Chang-Refsdal case). 
On the left side, configurations with $d^{\,[1]} > 1$ are shown, while the right
side shows the dual configurations with the inverse $d^{\,[1]} < 1$. From the top
to the bottom, $d^{\,[1]} \to 1$ is approaced.}
\label{caustictransition}}
\end{figure*}

\subsection{Full lens equation}
Let us reconsider the lens equation as given by Eq.~(\ref{eq:label7}) and study
the vicinity of the secondary object by introducing new dimensionless
apparent position  
coordinates $(\hat x_1,
\hat x_2)$ and new dimensionless true position coordinates $(\hat y_1,\hat y_2)$.
The origin of 
the new dimensionless apparent position coordinates is at 
$(x_1,x_2) = (-d/2,0)$, namely at the position of the secondary object, and
the origin of the new true source position coordinates is at
$(y_1,y_2) = (-d/2 + m_1/d,0)$, namely at that true source position that has
an image at the position of the secondary object (see also
Di Stefano \& Mao (\cite{DisM})). Moreover, the new coordinates measure 
separations in units of
$\theta_\mathrm{E}^{\,[2]}$, so that
\begin{eqnarray}
\hat x_1  & = & \left(x_1+d/2\right)/{\sqrt{1-m_1}}\,, \nonumber \\
\hat x_2 & = & x_2/{\sqrt{1-m_1}}\,, \nonumber \\
\hat y_1 & = & \left(y_1+d/2-m_1/d\right)/{\sqrt{1-m_1}}\,, \nonumber \\ 
\hat y_2 & = & y_2/{\sqrt{1-m_1}}\,. 
\label{eq:coords2}
\end{eqnarray}

Using these definitions, the lens equation can be written in the new coordinates as 
\begin{eqnarray}
\hat y_1(\hat x_1,\hat x_2) & = & \hat x_1 - \frac{\hat x_1}{\hat x_1^2+\hat x_2^2}
- \frac{1}{\sqrt{q}}\frac{\sqrt{q} \hat x_1 - d^{\,[1]}}{(\sqrt{q} \hat x_1 -d^{\,[1]})^2 + q \hat x_2^2}
- \nonumber \\
& & -\,\frac{1}{\sqrt{q}}\,\frac{1}{d^{\,[1]}} \nonumber \\
 & = & 
 \hat x_1 - \frac{\hat x_1}{\hat x_1^2+\hat x_2^2} + \nonumber \\
& & \!\!\!\!\!+\,\left[\hat x_1 - \sqrt{q}\,\frac{\hat x_1^2 + \hat x_2^2}{d^{\,[1]}}\right]\,
\frac{1}{(\sqrt{q} \hat x_1 -d^{\,[1]})^2 + q \hat x_2^2}\,, \nonumber \\
\hat y_2(\hat x_1,\hat x_2) & = & \hat x_2 - \frac{\hat x_2}{\hat x_1^2+\hat x_2^2}
-\frac{\hat x_2}{(\sqrt{q} \hat x_1 -d^{\,[1]})^2 + q \hat x_2^2}\,.
\label{eq:lenseq2}
\end{eqnarray}
Note that this is the full lens equation; no approximation has yet been made.

For $q=0$, these expressions yield the lens equation for a Chang-Refsdal lens, i.e.
\begin{eqnarray}
\hat y_1(\hat x_1,\hat x_2) & = & (1+\gamma) \hat x_1 - \frac{\hat x_1}{\hat x_1^2 +
	\hat x_2^2}\,, \nonumber \\
\hat y_2(\hat x_1,\hat x_2) & = & (1-\gamma) \hat x_2 - \frac{\hat x_2}{\hat x_1^2 +
	\hat x_2^2}\,, 
\end{eqnarray}
with the shear parameter
\begin{equation}
\gamma = 1/[d^{\,[1]}]^2\,.
\end{equation}
Thus for small mass ratios $q$, the properties of the lens are determined 
mainly by the angular separation parameter
$d^{\,[1]}$ and depend only weakly on $q$.

\subsection{Taylor-expansion}
The connection of the binary lens equation with the Chang-Refsdal lens equation can also be seen
by expanding the deflection due to the primary object 
at the position of the secondary object.
Expressing Eq.~(\ref{eq:lenseq2}) in terms of $d^{\,[2]}$ instead of $d^{\,[1]}$, 
the lens equation reads
\begin{eqnarray}
\hat y_1(\hat x_1,\hat x_2) & = & \hat x_1 - \frac{\hat x_1}{\hat x_1^2+\hat x_2^2}
- \frac{1}{q}\,\frac{\hat x_1 - d^{\,[2]}}{(\hat x_1 -d^{\,[2]})^2 + \hat x_2^2}
- \nonumber \\
& & \quad 
-\, \frac{1}{q}\,\frac{1}{d^{\,[2]}}\,, \nonumber \\
\hat y_2(\hat x_1,\hat x_2) & = & \hat x_2 - \frac{\hat x_2}{\hat x_1^2+\hat x_2^2}
- \frac{1}{q}\,\frac{\hat x_2}{(\hat x_1 -d^{\,[2]})^2 + \hat x_2^2}\,.
\end{eqnarray}
The factor $1/q$ gives the strength of the deflection due to the primary object. 
The Taylor-expansion of
\begin{eqnarray}
\hat g_1(\hat x_1, \hat x_2)
& \equiv & \frac{\hat x_1 - d^{\,[2]}}{(\hat x_1-d^{\,[2]})^2 + \hat x_2^2}\,, 
\nonumber \\
\hat g_2(\hat x_1,\hat x_2) & \equiv & \frac{\hat x_2}{(\hat x_1-d^{\,[2]})^2 + \hat x_2^2}
\end{eqnarray}
around $(\hat x_1, \hat x_2) = (0,0)$ is derived in Appendix~A and
yields with $\hat d = d^{\,[2]} = d^{\,[1]}/\sqrt{q}$
\begin{eqnarray}
\hat y_1 (\hat x_1,\hat x_2) &=& \hat x_1 - \frac{\hat x_1}{\hat x_1^2 + \hat x_2^2}
- \frac{1}{\sqrt{q}\,d^{\,[1]}} 
+ \nonumber \\
 & & \!\!\!\!\!\!\!\!\!\!\!\!\!\!\!\!\!\!\!\!\!\!\!\!\!\!\!\!\!\!
 +\,\sum_{n=0}^{\infty}
 \frac{q^{\frac{n-1}{2}}}{\left[d^{\,[1]}\right]^{n+1}}
\sum_{k=0}^{\lfloor n/2\rfloor} {\textstyle{n \choose 2k}}\,\hat x_1^{n-2k}\,\hat x_2^{2k}\,(-1)^k 
\,, \nonumber \\
\hat y_2 (\hat x_1,\hat x_2) & = & \hat x_2 - \frac{\hat x_2}{\hat x_1^2 + \hat x_2^2} - \nonumber \\
 & & \!\!\!\!\!\!\!\!\!\!\!\!\!\!\!\!
 \!\!\!\!\!\!\!\!\!\!\!\!\!\!-\,\sum_{n=1}^{\infty} \frac{q^{\frac{n-1}{2}}}{\left[d^{\,[1]}\right]^{n+1}}
\sum_{k=0}^{\lfloor \frac{n-1}{2}\rfloor} {\textstyle{n \choose 2k+1}}\,\hat x_1^{n-2k-1}\,\hat x_2^{2k+1}\,(-1)^k \,. 
\end{eqnarray}
As shown in Appendix~B, these series converge for
$\sqrt{\hat x_1^2+\hat x_2^2} < d^{\,[2]}$, i.e.\ within a circle around the
secondary object that has the primary object on its circumference. Note that this is
the maximal possible region of convergence, because the functions
$\hat y_1(\hat x_1,\hat x_2)$ and $\hat y_2(\hat x_1,\hat x_2)$ themselves
diverge for $(\hat x_1,\hat x_2) = (d^{\,[2]},0)$.
Taking into account terms up to $n = 4$, the lens equation reads
\begin{eqnarray}
\hat y_1 (\hat x_1,\hat x_2) &=& \hat x_1 - \frac{\hat x_1}{\hat x_1^2 + \hat x_2^2} +
\frac{1}{[d^{\,[1]}]^2}\,\hat x_1 + \nonumber \\
& & +\,\frac{\sqrt{q}}{[d^{\,[1]}]^3}\,(\hat x_1^2 - \hat x_2^2)
+ \frac{q}{[d^{\,[1]}]^4}\,(\hat x_1^3 - 3 \hat x_1 \hat x_2^2)\,, \nonumber \\
\hat y_2 (\hat x_1,\hat x_2) &=& \hat x_2 - \frac{\hat x_2}{\hat x_1^2 + \hat x_2^2} -
\frac{1}{[d^{\,[1]}]^2}\,\hat x_2 - \nonumber \\
& & -\,\frac{\sqrt{q}}{[d^{\,[1]}]^3}\, 2 \hat x_1 \hat x_2
- \frac{q}{[d^{\,[1]}]^4}\,(3 \hat x_1^2 \hat x_2 - \hat x_2^3)\,.
\end{eqnarray}
One sees that the Chang-Refsdal lens equation is revealed as the Taylor-expansion up to $n=1$.
The term for $n=0$ gives the shift of $1/(\sqrt{q}\,d^{\,[1]})$ in the $\hat y_1$-coordinate.
If one keeps the distance $d^{\,[1]}$ fixed, every higher-order term involves an additional factor
of $\sqrt{q}/d^{\,[1]}$, so that the series converges faster for smaller $q$ and the Chang-Refsdal lens
equation is the exact lens equation for $q \to 0$.

To see why the Chang-Refsdal approximation fails for $d^{\,[1]} \to 1$, let us consider the cases
$d^{\,[1]} > 1$ and $d^{\,[1]} < 1$ separately.

For $d^{\,[1]} > 1$, one has $\gamma < 1$. For
large $d^{\,[1]} \gg 1$, $\gamma$ tends to zero, and the critical curve tends to a circle with radius 1
($|\vec{\hat x}| \sim 1$), so that the criterion for the convergence of the
series at a multiple $\beta > 1$ of the critical curve distance 
becomes $d^{\,[1]} > \beta\sqrt{q}$, which can be fulfilled with a sufficiently large $d^{\,[1]}$ 
for any $0 < q \leq 1$. This means
that the Chang-Refsdal approximation is valid for suffiently wide binaries, 
where 'sufficiently wide'
means a smaller $d^{\,[1]}$ for smaller $q$. As $d^{\,[1]} > 1$ tends to 1, the critical curve grows like
$|1-\gamma|^{-1/2}$, so that at some point the convergence criterion can
no longer be fulfilled; this occurs 
closer to $d^{\,[1]} = 1$ for smaller $q$.

The case $d^{\,[1]} < 1$ corresponds to $\gamma > 1$, and, as for the other case, the critical curves grow
towards $d^{\,[1]} = 1$ like $|1-\gamma|^{-1/2}$. 
The Chang-Refsdal approximation thus also breaks down
as one approaches $d^{\,[1]} = 1$ from the side $d^{\,[1]} < 1$. For small $d^{\,[1]}$, i.e. large $\gamma$, the
position of the critical curves moves to $1/\sqrt{\gamma} = d^{\,[1]}$ above or below the $x_1$-axis,  and the
convergence criterion at a multiple of
this position becomes $q < 1/\beta^2$, so that for close binary systems
the Chang-Refsdal approximation is the better 
the smaller $q$, and fails for $q \to 1$, which also implies that the Chang-Refsdal
approximation fails in a close system near the primary object.

\subsection{Chang-Refsdal limit and caustics}
The way in which the caustics approach the Chang-Refsdal limit for $q \to 0$ for
different separations $d^{\,[1]}$ is illustrated
in Fig.~\ref{caustictransition}, using the
separations $d^{\,[1]} = 2$, 1.118, 1.06 (corresponding to the
shear parameters $\gamma = 0.25$, 0.8, 0.95) and the 
dual cases\footnote{See Sect.~\ref{sec:perturbation} for a discussion of this duality.}
$d^{\,[1]} = 0.5$, 0.894, 0.975 (corresponding to
$\gamma = 4$, 1.25, 1.053), and the mass ratios $q = 1$,
0.1, 0.01, $10^{-4}$, and 0. For
$d^{\,[1]} > 1$ ($\gamma < 1$) one sees that the Chang-Refsdal caustic evolves as the
left part of the six-cusp caustic for $q = 1$. The transition between the six-cusp caustic
and the two four-cusp caustics occur for $d^{\,[1]} = 2, 1.118, 1.026$ at
$q_\mathrm{w} = 0.203$, $4.6\cdot 10^{-4}$, and $5.1\cdot 10^{-6}$ respectively.
For $d^{\,[1]} > 2\,\sqrt{2}$, one has two four-cusp caustics (i.e.\ the wide binary
case) for all mass ratios. 
For $d^{\,[1]} < 1$ ($\gamma > 1$), the two triangular shaped Chang-Refsdal caustics
evolve from the two triangular shaped caustics for $q = 1$. If
$d^{\,[1]} \geq (3/4)^{3/4}$, there is a transition from close to intermediate binary
and back to close binary, while for $d^{\,[1]} < (3/4)^{3/4}$, one has a close binary
for any mass ratio. While $d^{\,[1]} < (3/4)^{3/4} \approx 0.806$ is fulfilled for
$d^{\,[1]} = 0.5$, the two other cases show transitions at
$d_\mathrm{c} = 0.581$ and $d_\mathrm{c} = 4.5\cdot 10^{-3}$ ($d^{\,[1]} = 0.894$) or
$d_\mathrm{c} = 0.899$ and $d_\mathrm{c} = 4.2\cdot 10^{-5}$ ($d^{\,[1]} = 0.975$).
Both for $d^{\,[1]} < 1$ and
for $d^{\,[1]} > 1$ one sees that the Chang-Refsdal limit is approached more slowly
as $d^{\,[1]} \to 1$.

\section{The perturbative picture for small mass ratios and its limits}
\label{sec:perturbation}
\begin{figure*}
\resizebox{12cm}{!}{\includegraphics{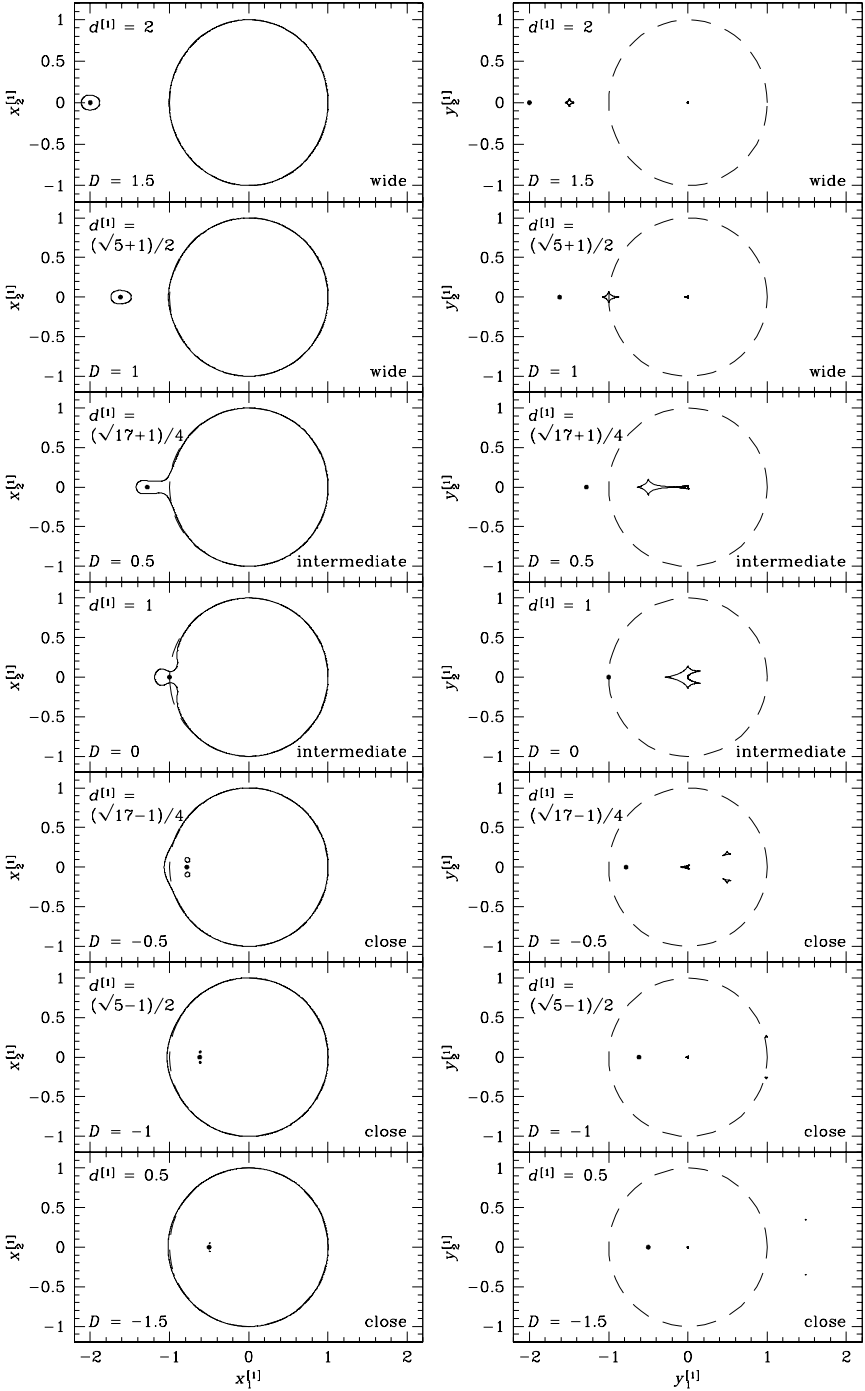}}
\hfill
\parbox[b]{55mm}{
\caption{Positions and shapes of the critical curves (left) and  caustics (right)
for several separations between the
lens objects and a fixed mass ratio $q = 0.01$. The coordinates are centered
on the position of the primary object and in units of the angular Einstein
radius $\theta_\mathrm{E}^{\,[1]}$ of the primary object. The Einstein ring is
shown as a dashed line. The secondary object is located to the left of the
primary at
a distance $d^{\,[1]}$, its position is indicated by a black dot. The 'center'
of the caustic near the secondary object is located at $D = d^{\,[1]}-1/d^{\,[1]}$ to
the left of the
primary object for $d^{\,[1]} > 1$, at the primary object for $d^{\,[1]} = 1$, and
at $-D$ to the right of the primary object for $d^{\,[1]} < 1$.}
\label{configs}}
\end{figure*}

The idea of the perturbative picture (Gould \& Loeb~\cite{GL}; Gaudi \& Gould~\cite{GG};
Griest \& Safizadeh~\cite{GS}) is that the behaviour of the binary lens for small ratios is
mainly determined by the primary object and that the secondary object perturbs one of the two
images due to lensing by the primary object.
For the lensing by the primary object, one has the lens equation
\begin{equation}
\vec y^{\,[1]} = \vec x^{\,[1]} - \frac{\vec x^{\,[1]}}{|\vec x^{\,[1]}|^2}\,,
\label{eq:lenseqpt}
\end{equation}
where the coordinates are centered on the primary object and denote angular coordinates in
units of $\theta_\mathrm{E}^{\,[1]}$. In polar coordinates, this means that the radii
fulfill the equation
\begin{equation}
y^{\,[1]} = x^{\,[1]} - \frac{1}{x^{\,[1]}}
\end{equation}
and that the polar angle is conserved.
Thus, for a given position $y^{\,[1]} > 0$, one obtains two images
\begin{equation}
x_{\pm}^{\,[1]} = \frac{1}{2}\left(y^{\,[1]}\pm\sqrt{{y^{\,[1]}}^2 + 4}\right)\,,
\end{equation}
where $x_{+}^{\,[1]}$ denotes the 'major image', and $x_{-}^{\,[1]}$ denotes the 'minor image',
and
\begin{equation}
x_{+}^{\,[1]} x_{-}^{\,[1]} = -1\,.
\end{equation}
Since
\begin{equation}
x_{+}^{\,[1]} \geq 1\,, \quad
-1 \leq x_{-}^{\,[1]} < 0 \,,
\end{equation}
the major image is always located outside the Einstein ring, while the minor image is always
located inside the Einstein ring. The total magnification of the two images is
\begin{equation}
\mu_\mathrm{tot} = \frac{{y^{\,[1]}}^2 + 2}
{y^{\,[1]}\,\sqrt{{y^{\,[1]}}^2 + 4}}\,,
\end{equation}
and the magnification of the individual images is
\begin{eqnarray}
\mu(x_{+}^{\,[1]}) & = &\frac{1}{2}\,(\mu_\mathrm{tot}+1) > 1\,,\nonumber \\
\mu(x_{-}^{\,[1]}) & = & \frac{1}{2}\,(\mu_\mathrm{tot}-1) =
\mu(x_{+}^{\,[1]}) -1
\,,
\end{eqnarray}
so that the major image is the brighter and the minor image is the dimmer image.
It has been shown in Sect.~\ref{sec:vicsec} that to
achieve the Chang-Refsdal lens limit near a secondary object located at $d^{\,[1]}$ from the primary
object, the origin of the true source position coordinates is not at the position of the secondary object
but shifted
by $1/d^{\,[1]}$ towards the primary object, so that it is located at
\begin{equation}
D = d^{\,[1]} - 1/d^{\,[1]}\,,
\label{eq:cceq}
\end{equation}
which can also be understood as the secondary object being located at the position of an image of a source object at $D$  
under lensing by the primary object, as can be seen
by setting $x^{\,[1]} = d^{\,[1]}$ and $y^{\,[1]} = D$ in
Eq.~(\ref{eq:lenseqpt}). The coordinate $D$ is identical to the center of the Chang-Refsdal-like
caustic corresponding to the critical curves near the secondary object.

Identifying Eq.~(\ref{eq:cceq}) with the lens equation of a point lens, one sees that there are 
two solutions for the position of the secondary object that put a caustic at $D$, namely
\begin{equation}
d_{\pm}^{\,[1]} = \frac{1}{2}\left(D \pm \sqrt{D^2+4}\right)\,,
\end{equation}
where $d_{+}^{\,[1]}$ corresponds to the major image, while 
$d_{-}^{\,[1]}$ corresponds to the minor image. A source object close to the caustic therefore has one image at the position of the secondary object, where this image is the
the major image
for $d^{\,[1]} = d_{+}^{\,[1]} > 1$ and the minor image for $|d^{\,[1]}| = |d_{-}^{\,[1]}| < 1$.
The secondary object thus perturbs the major image for $d^{\,[1]} > 1$, and the 
minor image for $|d^{\,[1]}| < 1$. As can be seen easily from Eq.~(\ref{eq:cceq}),
'dual' configurations
with
\begin{equation}
d^{\,[1]} \leftrightarrow -\frac{1}{d^{\,[1]}}
\end{equation}
yield perturbations at the same position $D$.
Since the shear in the Chang-Refsdal limit is $\gamma = 1/[d^{\,[1]}]^2$ (Sect.~\ref{sec:vicsec}),
the major image case corresponds to $\gamma > 1$ and therefore yields a
diamond-shaped caustic, while the minor image case corresponds to
$\gamma < 1$ and therefore yields two triangular-shaped caustics. 
A configuration with shear $\gamma$ has a dual configuration with shear $1/\gamma$.
In the major image case, the secondary object and the caustic are on the same side of the 
primary object, while they are on different sides in the minor image case.
To have a positive separation $d^{\,[1]}$, one has to choose $D \leq 0$ in the minor image case and 
the duality then reads
\begin{equation}
d^{\,[1]} \leftrightarrow \frac{1}{d^{\,[1]}} \quad \mbox{and} \quad D \leftrightarrow -D\,.
\end{equation}

The perturbative picture means that one of the two images of the point-mass lens (primary object) is 
disturbed by the secondary object. Since the secondary object acts as Chang-Refsdal lens (with pure shear)
on either the major
or the minor image, this image is split into two images or four images for both the
major image and the minor image case,
depending whether the
source is inside or outside the Chang-Refsdal caustic. Therefore, the 3 or 5 images of the binary lens
are the 2 or 4 images of the Chang-Refsdal lens of the affected primary lens image, and the unaffected
image. It is interesting to note that the Chang-Refsdal lens has been invented to study the influence of a 
single star in a galaxy that lenses a quasar (Chang \& Refsdal~\cite{CRlens}, \cite{CR2}), where the 
star splits one of the (macro-)images due to the galaxy into microimages.
In the case of binary lenses with small mass ratios, one has a similar situation,
where the primary object takes the function of the galaxy and the secondary object takes the 
function of the star.

For a given mass ratio $q = 0.01$, the critical curves and caustics and the positions of the
two lens objects are shown in Fig.~\ref{configs}. 
The position of the secondary object is shown by a small dot and the primary object
is located at the coordinate origin; distances are measured in units of 
$\theta_\mathrm{E}^{\,[1]}$.
The secondary object is always located to the left of the primary object. The center
of the caustic of the secondary object is located to the left of the primary object for $d^{\,[1]} > 1$, to the right for $d^{\,[1]} < 1$, and at the 
 primary for $d^{\,[1]} = 1$. Note the growth of the caustics for
$d^{\,[1]} \to 1$, and the transition from wide binaries to intermediate 
binaries (at $d_\mathrm{w}^{\,[1]} \approx 1.340$) 
to close binaries (at $d_\mathrm{c}^{\,[1]} \approx 0.870$). The
center of the caustic of the secondary object is inside the Einstein ring of the primary
object for $(\sqrt{5}-1)/2 < d^{\,[1]} < (\sqrt{5}+1)/2$. Since planets in this region
are favoured to be detected in current searches
(Gould \& Loeb~\cite{GL}; Bolatto \& Falco~\cite{BF}; Bennett \& Rhie~\cite{BR};
Gaudi \& Gould~\cite{GG}; Wambsganss~\cite{Wambs}; Griest \& Safizadeh~\cite{GS}),
this region has been
called 'lensing zone'. 

One sees that there is a finite region around $d^{\,[1]} = 1$ in which the binary lens cannot be approximated by a point-mass
lens at the primary object plus a local distortion of one image by the secondary object described by 
a Chang-Refsdal lens.

In fact, the Chang-Refsdal approximation supports wide binary lenses, where $\gamma < 1$, and
close binary lenses (near the secondary object), 
where $\gamma > 1$, while the 3rd topology of intermediate binary lenses has
no counterpart by means of the Chang-Refsdal approximation.
The perturbative picture therefore fails
for  
$d_\mathrm{c}^{\,[1]} < d^{\,[1]} < d_\mathrm{w}^{\,[1]}$.
However, for $q \to 0$, $d_\mathrm{c}^{\,[1]} \to 1$ and $d_\mathrm{w}^{\,[1]} \to 1$
so that the region for intermediate binaries vanishes, and 
for a given $d^{\,[1]} \neq 1$, the Chang-Refsdal approximation is valid for a 
sufficiently small mass ratio $q$, where 'sufficiently small' means
a smaller value as $d^{\,[1]} \to 1$.

\section{The central caustic}
\label{sec:centralcaust}

\subsection{Wide binaries and Taylor-expansion}
In Sects.~\ref{sec:vicsec} and~\ref{sec:perturbation}, the effect of the primary object on the secondary object has been studied. However, since there
is an inherent symmetry upon the interchange of the primary with the secondary object,
the secondary object also influences the vicinity of the primary object,
which leads to the existence of a small 
'central caustic'\footnote{as it has been named by Griest \& Safizadeh~(\cite{GS})}, which is located
at the position of the primary object shifted by the deflection due to the secondary object for wide binaries.
For close binaries, there is also a central caustic which is 
located at the center of mass.
Let us first investigate the
wide binary configuration.

To investigate the effect of the secondary on the primary object, one can choose coordinates in analogy to those used for the
discussion of 
the effect of the primary on the secondary object, given by Eq.~(\ref{eq:coords2}),
by just exchanging the roles of the two lens objects.
Let the coordinates 
$(\tilde x_1,
\tilde x_2)$ and $(\tilde y_1,\tilde y_2)$ measure 
angular positions in units of  
$\theta_\mathrm{E}^{\,[1]}$ and let the origin of 
the dimensionless apparent position coordinates be at the position of the primary object, i.e.\ at $(x_1,x_2) = (d/2,0)$.
The origin of
the dimensionless true position coordinates is chosen at 
$(y_1,y_2) = (d/2 - m_2/d,0)$. A source object at this position 
has an image at the position of the primary object due to the deflection of the secondary object.
This means that
\begin{eqnarray}
\tilde x_1 = \left(x_1-d/2\right)/{\sqrt{m_1}}\,,\quad 
\tilde x_2 = x_2/{\sqrt{m_1}}\,, \nonumber \\
\tilde y_1 = \left(y_1-d/2+m_2/d\right)/{\sqrt{m_1}}\,,\quad 
\tilde y_2 = y_2/{\sqrt{m_1}}\,. 
\end{eqnarray}
The lens equation in these coordinates reads
\begin{eqnarray}
\tilde y_1(\tilde x_1,\tilde x_2) & = & \tilde x_1 - \frac{\tilde x_1}{\tilde x_1^2+\tilde x_2^2}
- q\, \frac{\tilde x_1 + d^{\,[1]}}{(\tilde x_1 +d^{\,[1]})^2 + \tilde x_2^2}
+ \frac{q}{d^{\,[1]}}\,, \nonumber \\
\tilde y_2(\tilde x_1,\tilde x_2) & = & \tilde x_2 - \frac{\tilde x_2}{\tilde x_1^2+\tilde x_2^2}
- q\, \frac{\tilde x_2}{(\tilde x_1 +d^{\,[1]})^2 + \tilde x_2^2}\,.
\end{eqnarray}

The expansion of the deflection terms
\begin{eqnarray}
\tilde g_1(\tilde x_1, \tilde x_2)
& = & \frac{\tilde x_1 + d^{\,[1]}}{(\tilde x_1+d^{\,[1]})^2 + \tilde x_2^2}\,, 
\nonumber \\
\tilde g_2(\tilde x_1,\tilde x_2) & = & \frac{\tilde x_2}{(\tilde x_1+d^{\,[1]})^2 + \tilde x_2^2}
\end{eqnarray}
has been derived in Appendix~A and yields with $\hat d  = - d^{\,[1]}$
the lens equation as
\begin{eqnarray}
\tilde y_1 (\tilde x_1,\tilde x_2) &=& \tilde x_1 - \frac{\tilde x_1}{\tilde x_1^2 + \tilde x_2^2} 
+ \frac{q}{d^{\,[1]}}  \nonumber - \\
 & & \!\!\!\!\!\!\!\!\!\!\!\!\!\!\!\!\!\!\!\!\!\!\!\!\!\!\!\!\!\!
-\,q \sum_{n=0}^{\infty}
 \frac{(-1)^n}{\left[d^{\,[1]}\right]^{n+1}}
\sum_{k=0}^{\lfloor n/2\rfloor} {\textstyle{n \choose 2k}}\,\tilde x_1^{n-2k}\,\tilde x_2^{2k}\,(-1)^k\,, \nonumber \\ 
\tilde y_2 (\tilde x_1,\tilde x_2) & = & \tilde x_2 - \frac{\tilde x_2}{\tilde x_1^2 + \tilde x_2^2} - \nonumber \\
 & & \!\!\!\!\!\!\!\!\!\!\!\!\!\!\!\!\!\!\!\!\!\!\!\!\!\!\!\!\!\!
-\,q\sum_{n=1}^{\infty} \frac{(-1)^{n+1}}{\left[d^{\,[1]}\right]^{n+1}}
\sum_{k=0}^{\lfloor \frac{n-1}{2}\rfloor} {\textstyle{n \choose 2k+1}}\,\tilde x_1^{n-2k-1}\,\tilde x_2^{2k+1}\,(-1)^k \,. 
\end{eqnarray}
As shown in Appendix~B, these series converge for
$\sqrt{\tilde x_1^2+\tilde x_2^2} < d^{\,[1]}$, i.e.\ within a circle around the
primary object that has the secondary object on its circumference. 
As for the case of the effect of the primary object near the secondary object,
this is
the maximal possible region of convergence, since 
$\tilde y_1(\tilde x_1,\tilde x_2)$ and $\tilde y_2(\tilde x_1,\tilde x_2)$
diverge for $(\tilde x_1,\tilde x_2) = (-d^{\,[1]},0)$.
Including the terms up to $n = 4$ yields
\begin{eqnarray}
\tilde y_1 (\tilde x_1,\tilde x_2) &=& \tilde x_1 - \frac{\tilde x_1}{\tilde x_1^2 + \tilde x_2^2} + \nonumber \\
& & +\,q \left[
\frac{\tilde x_1}{[d^{\,[1)}]^2} + \frac{\tilde x_2^2 - \tilde x_1^2}{[d^{\,[1]}]^3}
+ \frac{\tilde x_1^3 - 3 \tilde x_1 \tilde x_2^2}
{[d^{\,[1]}]^4}\right]\,, \nonumber \\
\tilde y_2 (\tilde x_1,\tilde x_2) &=& \tilde x_2 - \frac{\tilde x_2}{\tilde x_1^2 + \tilde x_2^2} + \nonumber \\
& & +\,q \left[
\frac{-\tilde x_2}{[d^{\,[1]}]^2}+ \frac{2 \tilde x_1 \tilde x_2}{[d^{\,[1]}]^3}
+ \frac{\tilde x_2^3 - \tilde x_1^2 \tilde x_2}
{[d^{\,[1]}]^4}\right]\,,
\end{eqnarray}

This shows that the Chang-Refsdal term involves a shear of 
\begin{equation}
\tilde \gamma = \frac{q}{[d^{\,[1]}]^2} = \frac{1}{[d^{\,[2]}]^2}\,,
\end{equation}
compared to a shear of $\gamma = 1/[d^{\,[1]}]^2$ for the action of the primary object on the secondary object.
The coordinate scales differ however by a factor of $\sqrt{q}$ between the Einstein radii of the primary object
and of the secondary object, so that the caustic near the primary object is smaller than that
near the secondary object by a factor of $\sqrt{q}$ for wide binaries. Note that for constant
$d^{\,[1]}$, the size of the caustic near the secondary object decreases
with $\sqrt{q}$ (but the shear remains constant), 
while both the size of the caustic near the primary object and the shear 
decreases with $q$. 

To understand the validity of the Chang-Refsdal approximations for different $d^{\,[1]}$ and $q$, 
arguments similar to those in Sect.~\ref{sec:vicsec} can be used.

For large $d^{\,[1]}$, $\tilde \gamma$ tends to zero, and the critical curve tends to a unit circle.
To have convergence at a multiple $\beta > 1$ of the distance of the critical curve, one has
therefore to fulfill the criterion $d^{\,[1]} > \beta$, which can be fulfilled for sufficiently large 
$d^{\,[1]}$.
Contrary to the vicinity of the secondary object, the convergence criterion does not depend on $q$, so that at a given
$d^{\,[1]}$, one has the same deviation from the Chang-Refsdal limit for any mass ratio $q$.
Since for $d^{\,[1]} \to 1$, 
the size of the critical curves increases like $|1-\gamma|^{-1/2}$, the Chang-Refsdal limit fails at
some $d^{\,[1]} \sim 1$ (which does not depend strongly on $q$). 
For $d^{\,[1]} \to 0$, the critical curves are at $1/\sqrt{\tilde \gamma} = d^{\,[2]} = d^{\,[1]}/\sqrt{q}$ above or below
the $x_2$-axis, and the convergence criterion becomes $q > \beta^2$, which cannot be fulfilled for $0 < q \leq 1$.
As already pointed out in Sect.~\ref{sec:vicsec}, the vicinity of the primary object can therefore not be approximated
by a Chang-Refsdal lens in a close binary system.

\subsection{Close binaries and the duality}
\begin{figure*}
\resizebox{\hsize}{!}{\includegraphics{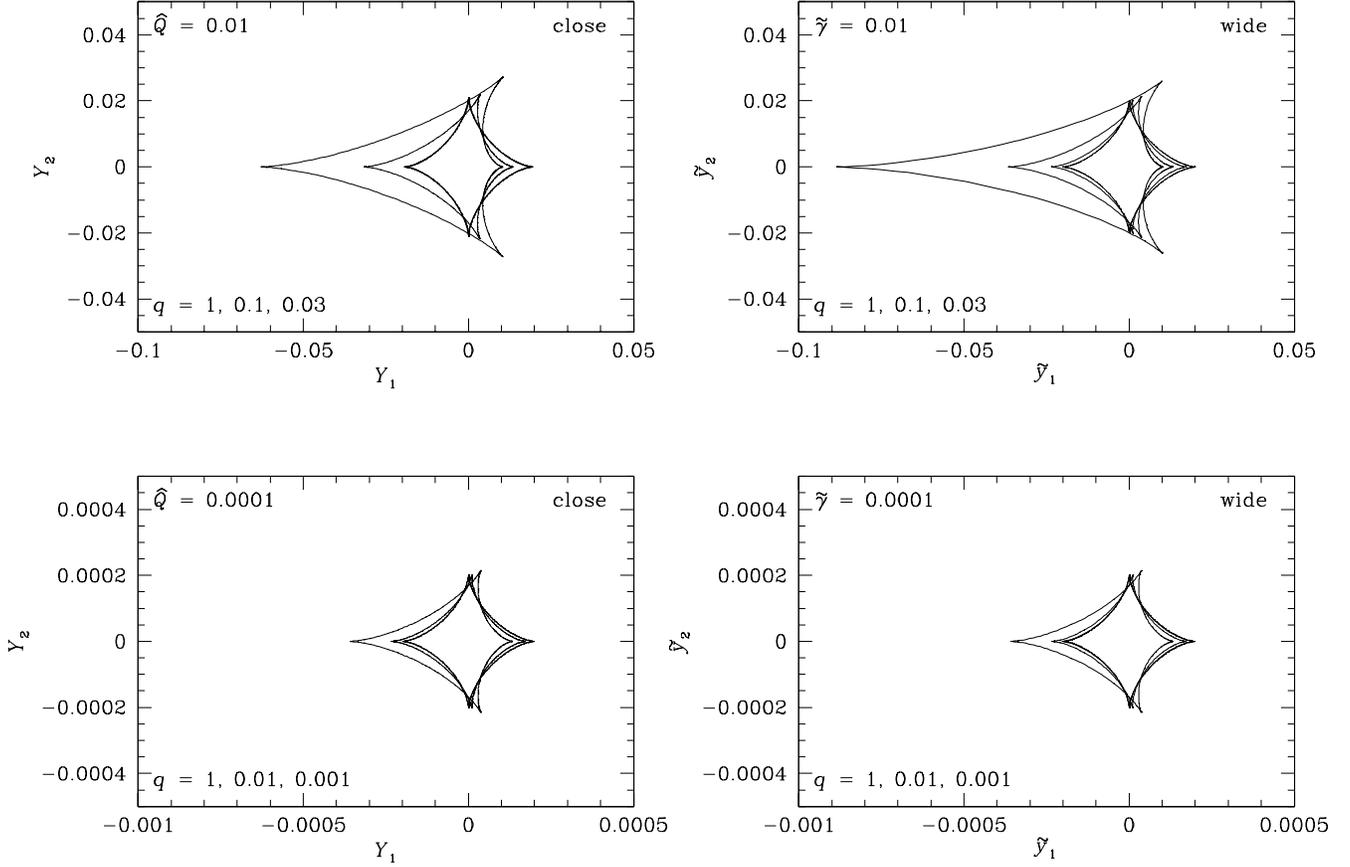}}
\caption{The central caustic for the binary lens for a given quadrupole moment $\hat Q$ (left) or a
given shear $\tilde \gamma$ (right) for different mass ratios $q$. 
The coordinates are centered on the center of mass for the close binary,
$\hat Q =\mathrm{const.}$ cases, and on the caustic (i.e. position of the primary shifted by deflection
of secondary object) for the wide binary, $\tilde \gamma = \mathrm{const.}$ cases.
The caustics for the corresponding quadrupole or Chang-Refsdal lenses are also
displayed; for the cases $\hat Q = 0.01$, $\hat Q = 0.0001$, and $\tilde
\gamma = 0.0001$, the difference between these caustics and the caustics
for a binary lens with $q=1$ cannot be seen. The asymmetry of the caustics increases towards smaller $q$. For
$q=1$, the close binaries have symmetric caustics.
For $\hat Q = 0.01$, the separations between the lens objects that
correspond to the mass ratios shown are $d^{\,[1]}$ = 0.283,
0.365, and 0.604.  
For $\tilde \gamma = 0.01$, the
corresponding separations are $d^{\,[1]} =$ 10, 3.16, 1.73. 
{\bf c} $\hat Q = 0.0001$.
For $\hat Q = 0.0001$, the corresponding separations are $d^{\,[1]} =$ 0.028, 0.102,  and 0.317, and for
$\tilde \gamma = 0.0001$, 
the corresponding separations are $d^{\,[1]} =$ 100, 10, and 3.2. 
}
\label{fig:qgconst}
\end{figure*}

\begin{figure*}
\resizebox{\hsize}{!}{\includegraphics{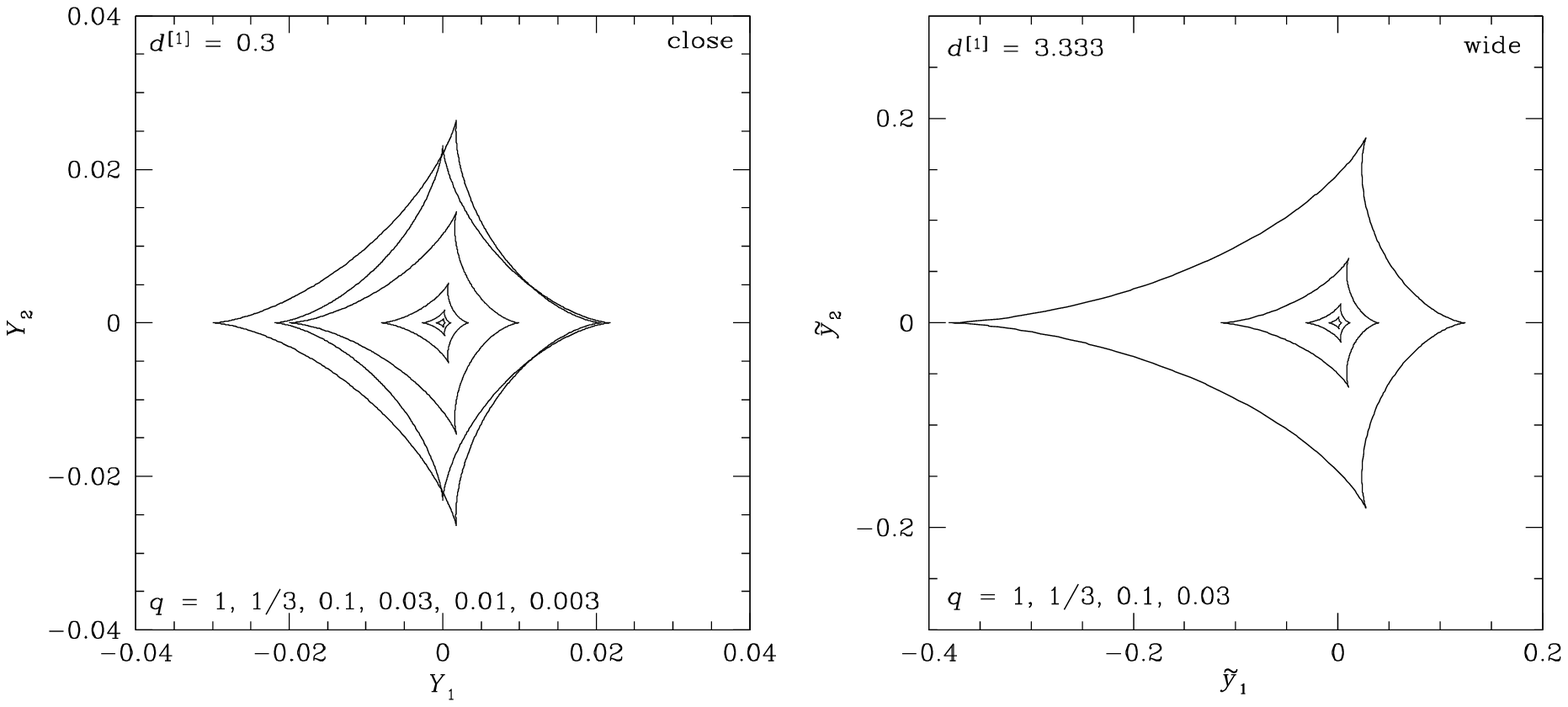}}
\caption{The central caustic for the binary lens for a given angular separation 
$d^{\,[1]}$ and various mass ratios $q$.
The coordinates are centered on the center of mass for the close binaries (left),
and on the caustic (i.e. position of the primary shifted by deflection
of secondary object) for the corresponding wide binaries (right).}
\label{fig:dconst}
\end{figure*}

In the previous sections, it has already been shown that for a wide binary lens, the vicinity of each of the
lens objects can be approximated by a pure-shear Chang-Refsdal lens with $\gamma < 1$
for sufficiently large separations. In addition, it has been shown that the
vicinity of the secondary object for small separations $d^{\,[1]} < 1$ can be approximated by a pure-shear
Chang-Refsdal lens with $\gamma > 1$, while the vicinity of the primary object cannot be described by
a Chang-Refsdal lens.

As pointed out in Sect.~\ref{sec:quad}, for distances from the lens that are much larger than the separation
between the two lens objects,
the lens equation can be approximated by means of the multipole expansion, where the
absolute value of the eigenvalues of the quadrupole moment $\hat Q$ is given
by Eq.~(\ref{eq:quadmoment}).
The main effect of the binary lens in this limit
is that of a point-mass lens located at the center of mass with the
total mass of both lens objects and a distortion by the quadrupole term, which, for $\hat Q < \frac{1}{12}$,
yields to a diamond-shaped caustic at the origin with size $4 \hat Q$, and a critical curve close to the unit circle. 
For $d^{\,[1]} \ll 1$, sources close to the coordinate origin are mapped to image positions near the critical curves, for 
which the distance from the center of mass $|\vec X|$ is much larger
than the separation between the lens objects $d$. 
The quadrupole limit becomes invalid however for image positions $|\vec X| \sim d$,
especially in the vicinity of the secondary object for $q \ll 1$, where the Chang-Refsdal approximation becomes valid.

Though the lens equation cannot be described by a Chang-Refsdal lens, there are some similarities between the caustics.
A wide binary lens with a shear $\gamma$ at one of its lens objects and a close binary lens with $\hat Q = \gamma$
have similar caustics and
magnification patterns and can therefore yield ambiguous solutions for an observed binary lens
event. 

If one compares the quadrupole moment, Eq.~(\ref{eq:quadmoment}), with the shear $\tilde \gamma = q/[d^{\,[1]}]^2$ 
acting on the
primary object for wide binaries, one sees that these configurations correspond to dual cases
$d^{\,[1]} \leftrightarrow (1+q)^{3/2}/d^{\,[1]}$ or
$d \leftrightarrow (1+q)^{1/2}/d$. In the limit $q \ll 1$, one obtains approximately the same duality as
for the vicinity of the secondary object $d^{\,[1]} \leftrightarrow 1/d^{\,[1]}$. 
Moreover, for $q \sim 1$, the caustic for a close binary is symmetric and similar to the Chang-Refsdal limit
near the secondary object for a wide binary, which is degenerate in $q$ for small $q$.
Thus an 
ambiguity exists between about equal mass close binaries and small mass ratio wide binaries if only the vicinity of
the secondary object and no effect from the primary object is observed.
An example for this are the binary lens fits for MACHO LMC-1 performed by Dominik \& Hirshfeld (\cite{DoHi2}):
wide binary fits  with $\gamma$ = 0.043, 0.050, 0.051, 0.057, and close binary fits
$\hat Q$ = 0.041 or 0.047 were found all to be good fits to the data.

Griest \& Safizadeh (\cite{GS}) have investigated the asymmetry of the central caustic.
They find that it
has a 'tip' towards the secondary object located at
\begin{equation}
\sigma_\mathrm{t} = \frac{q d^{\,[1]}}{\left(1-d^{\,[1]}\right)^2}
\end{equation}
from the primary object
for small $q$ and $d^{\,[1]}$ not close to $d^{\,[1]} = 1$. They also note that 
$\sigma_\mathrm{t}$ is
invariant under the duality transformation $d^{\,[1]} \leftrightarrow 1/d^{\,[1]}$,
so that similar caustics appear in both cases.
However, this can only be decided after moving to the adequate coordinate system, which is the 
center-of-mass in the case of the close binary, and the position of the primary corrected by the deflection by the
secondary object (this is where the caustic is centered) in the case of a wide binary.
The location of the tip in the center-of-mass system is (for $q \ll 1$)
\begin{eqnarray}
s_\mathrm{t} & = & \sigma_\mathrm{t} - q\,d^{\,[1]} \nonumber \\
& = & \left[\frac{1}{\left(1-d^{\,[1]}\right)^2} - 1\right]\,q\,d^{\,[1]}\,.
\end{eqnarray}
For $d^{\,[1]} \ll 1$, one obtains
\begin{eqnarray}
s_\mathrm{t} & \simeq & 2\,q\,[d^{\,[1]}]^2\left(1+\frac{3}{2} d^{\,[1]} + 2 [d^{\,[1]}]^2\right) \nonumber \\
& = & 2 \hat Q \left(1+\frac{3}{2} \sqrt{\hat Q/q} + 2 \hat Q/q\right)\,,
\label{eq:asym1}
\end{eqnarray}
so that $s_\mathrm{t}$ tends to the value $2 \hat Q$ for $\hat Q/q \ll 1$, i.e. $d^{\,[1]} \ll 1$, 
as expected for a quadrupole lens for
$\hat Q \ll 1$.

\begin{figure}[t]
\resizebox{\hsize}{!}{\includegraphics{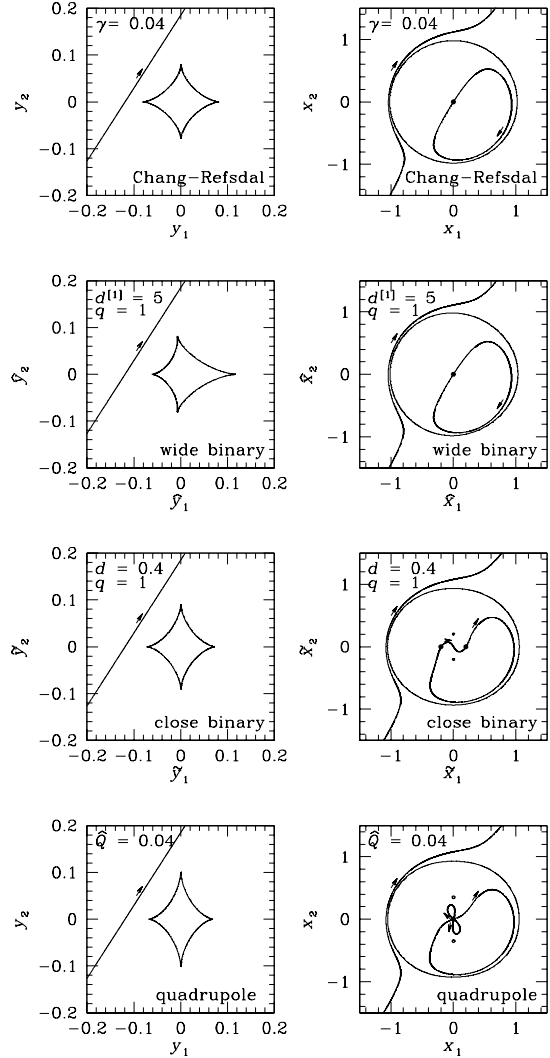}}
\caption{Diamond-shaped caustics and source trajectory (left), 
critical curves and image trajectories (right) 
for 4 different lens
models yielding similar caustics. 
Arrows indicate the direction of motion, small filled 
circles separate the image tracks and indicate images for $t \to \pm \infty$. 
The source passes at $u_0 = 0.1$ from the center of the caustic (coordinate center), and the
angle between $y_1$-axis and the source trajectory is $\alpha = 1~\mbox{rad}$.
For the wide binary model, the critical curve and caustic near the left lens object
are shown. 
There are 
mirror-symmetric critical curves and caustics near the other lens object as well as
another image track consisting of images
with small magnifications near the other lens object which are not shown.
For the close binary, the two small triangular shaped caustics and the
coresponding small nearly circular critical curves along the $x_2$-axis 
are not shown.
The positions of the filled circles coincide with the positions of the two lens objects. 
For the quadrupole lens,
there are two small triangular caustics along the $y_2$-axis and
two corresponding small nearly circular critical curves near the
$x_2$-axis which are not shown.}
\label{fig:imtracks}
\end{figure}

For a wide binary, the location of the tip relative to the caustic center is (for $q \ll 1$) 
\begin{eqnarray}
S_\mathrm{t} & = & \sigma_\mathrm{t} - \frac{q}{d^{\,[1]}} \nonumber \\
& = & \left[\frac{1}{\left(1-1/d^{\,[1]}\right)^2} - 1\right]\,\frac{q}{d^{\,[1]}}\,.
\end{eqnarray}
For $d^{\,[1]} \gg 1$, one obtains
\begin{eqnarray}
S_\mathrm{t} & \simeq & 2\,\frac{q}{[d^{\,[1]}]^2}\left(1+\frac{3}{2\,d^{\,[1]}} + 
\frac{2}{[d^{\,[1]}]^2}\right) \nonumber \\
& = & 2 \tilde \gamma \left(1+\frac{3}{2} \sqrt{\tilde \gamma/q} + 2 \tilde \gamma/q\right)\,,
\label{eq:asym2}
\end{eqnarray}
so that $S_\mathrm{t}$ tends to the value $2 \tilde \gamma$ for $\tilde \gamma/q \ll 1$, i.e. $d^{\,[1]} \gg 1$, 
as expected for a Chang-Refsdal lens for $\tilde \gamma \ll 1$.
In fact, since one just subtracts the dual coordinate origin corrections,
$s_\mathrm{t}(d^{\,[1]}) = S_\mathrm{t}(1/d^{\,[1]})$ and 
$s_\mathrm{t}(\hat Q) = S_\mathrm{t}(\tilde \gamma)$, 
the dual cases yield very similar caustics.

In both cases, for given $\hat Q$ or given $\tilde \gamma$, the limit and therefore the symmetry of the 
diamond-shaped caustic is approached faster for larger $q$, because this corresponds to a separation
further from $d^{\,[1]} = 1$, as illustrated in Fig.~\ref{fig:qgconst}.
While the size of the central caustic in $y_2$-
direction does not
grow much as $d^{\,[1]} \to 1$, an elongation towards the secondary object evolves
for small $q$ yielding to the tip described above.

However, Eqs.~(\ref{eq:asym1}) and~(\ref{eq:asym2}) show that
the relative asymmetry for $q \ll 1$ depends only on $d^{\,[1]}$, so that for the
same $d^{\,[1]}$ there is no difference among different mass ratios (see Fig.~\ref{fig:dconst}).
While for close binaries, the caustic becomes symmetric as $q \to 1$, because of the symmetry of the lens with respect to
the center of mass, the asymmetry remains for wide binaries.

For a given $\hat Q$, there is a limit on the possible mass ratios $q$ arising
from the condition $d^{\,[1]} < 1$. For $q \ll 1$, this limit becomes
$q > \hat Q$. In the same way, for a given $\tilde \gamma$, there is a limit on
$q$ arising from $d^{\,[1]} > 1$, which becomes $q > \tilde \gamma$ for $q \ll 1$.

In both cases, the size of the caustic depends mainly on $\tilde \gamma$ or $\hat Q$, 
while the separation $d^{\,[1]}$ 
determines the asymmetry.

\section{Diamond-shaped caustics}
\label{sec:sim}

As argued in the previous sections, there are rough similarities and small differences between
the binary lens and its limiting cases, the quadrupole lens and the pure shear Chang-Refsdal lens.
In addition, there is a similarity between the quadrupole lens and the pure shear Chang-Refsdal
lens for $\gamma < 1$ if the shear and the absolute value of an eigenvalue 
of the quadrupole moment coincide.

In particular, there are several cases, where a diamond shaped caustic arises. This is the case
for both close and wide binary lenses, where in the first case there is one central caustic, while
in the second case, two diamond-shaped caustics near both lens objects occur. Both the pure shear
Chang-Refsdal lens and the quadrupole lens produce a diamond-shaped caustic, namely for
$\gamma < 1$ or $\hat Q < \frac{1}{12}$.

However, there is a difference concerning the number of images. Depending on whether the source is
inside or outside the caustic, the binary lens produces 3 or 5 images, while the Chang-Refsdal lens
produces 2 or 4 images, and the quadrupole lens produces 4 or 6 images. While at first sight this
seems to be an important difference, this difference is not so important if looking at the total
magnification because for a source close to the caustic, the images common to all lens models
have much larger magnification than the images that occur only for certain lens models.

In Fig.~\ref{fig:imtracks}, the image trajectories for a source passing close to the caustic
are shown for both close and wide equal mass binary lenses and the corresponding
quadrupole and Chang-Refsdal lenses. For the close binary lens the angular separation is
$d = 0.4$, while it is $d = 5$ for the wide binary lens. For the 
quadrupole lens, $\hat Q = 0.04$, and for the Chang-Refsdal lens  $\gamma = 0.04$.
The sign of $\gamma$ is chosen to be that one that arises from
the limit of a wide binary, so that the extent of the caustic along the $x_1$-axis is larger
than the extent along the $x_2$-axis. A close binary however has a caustic that has longer
extent along the $x_1$-axis.
The motion of the images induced by the motion of the source is indicated by arrows. Small
filled circles separate the different image tracks and mark image positions for $t \to \pm \infty$.

\section{Summary}
\label{sec:summary}
The critical curves and caustics of a binary lens have 3 different topologies, which for
any given mass ratio $q$ depend on $d^{\,[1]}$ which is the
angular separation between the lens objects in units of the angular Einstein radius
$\theta_{\rm E}^{\,[1]}$ of the heavier, 'primary', lens
object. These
3 different topologies can be categorized as 'close' (small $d^{\,[1]}$), 'intermediate' (around
$d^{\,[1]} = 1$), and 'wide' (large $d^{\,[1]}$). There are no close binaries for $d^{\,[1]} > 1$ and there
are no wide binaries for $d^{\,[1]} < 1$. The region of intermediate binaries vanishes for
$q \to 0$ like $q^{1/3}$. 

For close and wide binaries, the binary lens can be approximated by simpler models
if one considers
the vicinity of the caustics that occur near the primary or the secondary object.
For intermediate binaries, there is only one caustic which cannot be approximated in a similar way.

Near the secondary (less-massive) object, the deflection due to the primary object can be 
Taylor-expanded, where the Chang-Refsdal approximation corresponds to a truncation of this series.
The shear is given by 
$\gamma = 1/[d^{\,[1]}]^2$, so that for a wide binary lens, where $d^{\,[1]} > 1$, i.e. the secondary object being outside the Einstein
ring of the primary object, one has $\gamma < 1$, and therefore a diamond-shaped caustic. 
In the language of the perturbative
picture, the major image has been perturbed. 
For a close binary lens, $d^{\,[1]} < 1$, i.e.
the secondary object is inside the Einstein ring of the primary object, 
$\gamma > 1$ and there are two triangular shaped caustics, and the minor image is perturbed.
'Dual' configurations $d^{\,[1]} \leftrightarrow 1/d^{\,[1]}$ (i.e. $\gamma \leftrightarrow 1/\gamma$) yield
caustics at the same distance (but on different sides) from the primary object.

This shows how the 3 binary lens topologies are mapped onto the 2 pure shear Chang-Refsdal topologies
for $q \ll 1$, where the intermediate topology has no counterpart.

For a given $d^{\,[1]}$, the lens equation in the vicinity of the secondary object approaches the
Chang-Refsdal lens equation as $q \to 0$, i.e. the Chang-Refsdal approximation becomes better for smaller $q$.
The Chang-Refsdal approximation 
breaks down as
$d^{\,[1]} \to 1$, latest when the transition to an intermediate binary occurs. 
For smaller $q$,
$d^{\,[1]} = 1$ can be approached more closely without the approximation
breaking down.

In the vicinity of the primary object, the Taylor-expansion, and therefore the Chang-Refsdal
approximation, works only for wide binary lenses. The shear is
$\tilde \gamma = q/[d^{\,[1]}]^2$, and therefore the (central) caustic near the primary object
is smaller by $\sqrt{q}$ than that near the secondary object. 
For close binaries, the vicinity of the center of mass, which for $q \ll 1$ is located 
near the primary object,
can be described by means of multipole expansion. An approximate
model involves the monopole term, i.e. a point-lens with the total mass in the center of mass, and
the quadrupole term, with the absolute value of the eigenvalues of the quadrupole moment of
$\hat Q = q d^2/(1+q)^2 \simeq q [d^{\,[1]}]^2.$ The quadrupole lens and the pure shear Chang-Refsdal lens
have similar caustics and magnification patterns if $\hat Q = \tilde \gamma$.
In addition, the asymmetry
(due to higher order terms) is the same in both cases, so that there is an ambiguity between these cases for any given $q$.
The condition $\hat Q = \tilde \gamma$ implies the duality $d^{\,[1]} \leftrightarrow (1+q)^{3/2}/d^{\,[1]}$,
which for $q \ll 1$ is approximately the same duality as for the vicinity of the secondary object,
namely
$d^{\,[1]} \leftrightarrow 1/d^{\,[1]}$.
In fact, Albrow et al. (\cite{SMC2}) just found this duality in their discussion
of all possible models for the MACHO 98-SMC-1 event.
The relative asymmetry of the caustic depends only on $d^{\,[1]}$ with the exception of
$d^{\,[1]} \sim 1$ in the close binary case, since the caustic becomes
symmetric as $d^{\,[1]} \to 1$.
Therefore, $d^{\,[1]}$ determines the asymmetry, while 
$\tilde \gamma$ or $\hat Q$ determine the size of the caustic.

If one only observes the vicinity of the secondary object, an additional ambiguity 
enters due 
to the fact that one exhibits nearly a degeneracy in
$q$ for small $q$, which is also ambiguous to a close binary with nearly equal mass
(Dominik \& Hirshfeld~\cite{DoHi2}).

\begin{acknowledgements}
I would like to thank G. Covone, K. C. Sahu, and P. D. Sackett
for commenting on different versions of this manuscript.
\end{acknowledgements}

\begin{appendix}

\section{Taylor-expansion of the deflection term}
\label{app:real}
To expand deflection terms of the form 
\begin{eqnarray}
\hat g_1(\hat x_1, \hat x_2)
 & = & \frac{\hat x_1 - \hat d}{(\hat x_1-\hat d)^2 + \hat x_2^2}\,, \nonumber \\
\hat g_2(\hat x_1,\hat x_2)  & = & \frac{\hat x_2}{(\hat x_1-\hat d)^2 + \hat x_2^2}
\end{eqnarray}
around $(\hat x_1, \hat x_2) = (0,0)$, let us proceed as follows.
Let $g(x_1,x_2)$ be defined as
\begin{equation}
g(x_1,x_2) = \frac{1}{2}\,\ln(x_1^2+x_2^2)\,,
\end{equation}
and let indices to $g$ denote derivations with respect to $x_1$ or $x_2$. The first
derivatives are given by
\begin{equation}
g_1(x_1,x_2) = \frac{x_1}{x_1^2 + x_2^2}\,, \quad g_2(x_1,x_2) = \frac{x_2}{x_1^2 + x_2^2}\,.
\label{g1g2eq}
\end{equation}
Since $\hat g_i(\hat x_1, \hat x_2) = g_i(\hat x_1 - \hat d,\hat x_2)$, the expansion of
$\hat g_i$ around $(\hat x_1,\hat x_2) = (0,0)$ is equivalent to the expansion of
$g_i$ around $(x_1,x_2) = (-\hat d,0)$.

In Appendix~B, it is shown how the expansion for $g_1$ and $g_2$ (Eq.~(\ref{gexp})) 
can easily been obtained using complex
notation. However, the same result can also be obtained with two-dimensional real notation as shown here.

The next 3 derivatives of $g$ are given by
\begin{eqnarray}
g_{ij} & = & \frac{\delta_{ij}\,s - 2x_j x_j}{s^2}\,, \nonumber \\
g_{ijk}
& = & \frac{1}{s^3}\,\left[-2s(\delta_{ij} x_k + \delta_{jk} x_i + \delta_{ik} x_j) + 8 x_i x_j x_k\right]\,, \nonumber \\
g_{ijkl}
& = & \frac{1}{s^4}\,\left[-2s^2\,(\delta_{ij}\delta_{kl} + \delta_{jk}\delta_{il} + \delta_{ik}\delta_{jl})
\,+\right. \nonumber \\ & & +\,
8s\,(\delta_{ij} x_k x_l + \delta_{jk} x_i x_l + \delta_{ik} x_j x_l \,+ \nonumber \\ 
& & \quad \,+\left.\delta_{il} x_j x_k + \delta_{jl} x_i x_k + \delta_{kl} x_i x_j)
\,-\right. \nonumber \\
& & -\, \left.48 x_i x_j x_k x_l\right]\,, 
\end{eqnarray}
where
\begin{equation}
s = x_1^2 + x_2^2\,.
\end{equation}
In general, the expressions are symmetric in the indices and 
the $n$-th derivative involves a sum of up to $n$-th order polynomials divided by $s^{n}$.
For $(x_1, x_2) = (-\hat d,0)$, all terms involving $x_2$ vanish, the only remaining polynomials
are those $\propto x_1^{n}$, thus the $n$-th derivative at $(x_1,x_2) = (-\hat d,0)$ is
proportional to $1/\hat d^n$. 

Since $s$ is the scalar product of the vector $\vec x$ with itself, it transforms like a scalar under coordinate
transformation. Therefore, with $g = \frac{1}{2}\,\ln s$ being a scalar, the first
derivatives $g_i = x_i/s$ form a vector and
the higher derivatives (of order $n$) form
a symmetric (due to Schwarz's theorem) tensor of order $n$.  
Forming a symmetric tensor requires the terms in the enumerator to be of the form
\begin{equation}
\delta_{\alpha_1,\alpha_2}\ldots{}\delta_{\alpha_{r-1},\alpha_r}\;x_{\beta_1}\ldots{}x_{\beta_s}\,,
\end{equation}
where $\alpha_i$ and $\beta_i$ denote indices of $g$ and $r+s = n$.\footnote{Moreover, all permutations
must be present in order to form a symmmetric tensor.} Note that $\delta_{ij}$ is the metric tensor in euclidian
space.
If there is an odd number of indices being equal to 2, a term involving $x_2$ is present, or 
a $\delta_{12}$ occurs, so that for $x_2 = 0$ all derivatives with an odd number of derivations with respect to
$x_2$ vanish.

Let $G_{n,m}(x_1,x_2)$ denote the $n$-th derivative of $g$, where $g$ is derived $m$ times with respect to $x_2$ and
$n-m$ times with respect to $x_1$, i.e. 
\begin{equation}
G_{n,m}(x_1,x_2) = \left(\frac{\partial}{\partial x_1}\right)^{n-m}
\left(\frac{\partial}{\partial x_2}\right)^{m}\,g(x_1,x_2)\,.
\end{equation}
Since 
\begin{equation}
g_{11}(x_1,x_2) = -g_{22}(x_1,x_2)
\end{equation}
one gets
\begin{equation}
G_{n,m+2}(x_1,x_2) = - G_{n,m}(x_1,x_2)\,.
\end{equation}
Moreover, $G_{n,0}(x_0,0)$ is given by the derivatives of the one-dimensional function
$h(x) = \ln x$, i.e.
\begin{eqnarray}
G_{n,0}(x_0,0)  & = & \frac{1}{2} \left(\frac{\partial}{\partial x_1}\right)^n (\ln(x_1^2+x_2^2))(x_0,0) \nonumber \\ 
 & & \!\!\!\!\!\!\!\!\!\!\!\!\!\!= \left(\frac{\mathrm{d}}{\mathrm{d}x}\right)^n\,(\ln x)(x_0) = (-1)^{n+1}\,\frac{(n-1)!}{x_0^n}\,.
\end{eqnarray}
Combining all properties of $G_{n,m}(x_0,0)$, it can be written as 
\begin{eqnarray}
G_{n,2k+1}(x_0,0) & = & 0 \,, \nonumber \\
G_{n,2k}(x_0,0) & = & (-1)^k\,\frac{(-1)^n\,(n-1)!}{x_0^{n}}\,.
\label{Geq}
\end{eqnarray}
with $k \in \bbbn_0$.

Since the Taylor-expansion for a scalar function $f$ depending on two coordinates is given by
\begin{eqnarray}
f(\vec x_0 + \vec \varepsilon\,) & = & \sum_{n=0}^{\infty} \sum_{m=0}^{n} \frac{1}{(n-m)!\, m!}\,\cdot \nonumber \\
 & & \!\!\!\!\!\!\!\!\cdot\,\left(\frac{\partial}{\partial x_1}\right)^{n-m}
\left(\frac{\partial}{\partial x_2}\right)^{m}\,f(\vec x_0)\;\varepsilon_1^{n-m} \varepsilon_2^m\,,
\end{eqnarray}
the Taylor-expansions for $g_1$ and $g_2$ around $(x_0,0)$ read
\begin{eqnarray}
g_1(x_0+x_1,x_2) & = &
\sum_{n=0}^{\infty} \sum_{m=0}^{n} \frac{1}{(n-m)!\, m!}\,\cdot \nonumber \\
& & \cdot\,
G_{n+1,m}(x_0,0)\,
x_1^{n-m} x_2^m\,, \nonumber \\
g_2(x_0+x_1,x_2) & = & 
\sum_{n=0}^{\infty} \sum_{m=0}^{n} \frac{1}{(n-m)!\, m!}\,\cdot \nonumber \\
& & \cdot\,
G_{n+1,m+1}(x_0,0)\,
x_1^{n-m} x_2^m\,.
\end{eqnarray}

Inserting $G_{n,m}$ from Eq.~(\ref{Geq}) and omitting the terms with odd $m$ yields
\begin{eqnarray}
g_1(x_0+x_1,x_2) &=& 
\sum_{n=0}^{\infty}
\frac{(-1)^n}{x_0^{n+1}}
\sum_{k=0}^{\lfloor n/2\rfloor} {\textstyle{n \choose 2k}}\,x_1^{n-2k}\,x_2^{2k}\,(-1)^k 
\,,  \nonumber \\
g_2(x_0+x_1,x_2) & = & \nonumber \\
& & \!\!\!\!\!\!\!\!\!\!\!\!\!\!\!\!\!\!\!\!\!\!\!\!\!
\!\!\!\!\!\!\!\!\!\!\!\!\!\!\!\!\!\!\!\!
= \sum_{n=1}^{\infty} \frac{(-1)^{n+1}}{x_0^{n+1}}
\sum_{k=0}^{\lfloor \frac{n-1}{2}\rfloor} {\textstyle{n \choose 2k+1}}\,x_1^{n-2k-1}\,x_2^{2k+1}\,(-1)^k \,. 
\label{gexp}
\end{eqnarray}

With $x_0 = -\hat d$, one obtains the expansion of $\hat g_1$ and $\hat g_2$ around $
(\hat x_1, \hat x_2) = (0,0)$, which converges for $\sqrt{\hat x_1^2 + \hat x_2^2} < \hat d$
(see Appendix~B).

\section{Taylor-expansion of the deflection term in complex notation}
\label{app:complex}
As pointed out by Witt (1990), the lens equation can also be formulated by means
of one-dimensional complex coordinates instead of two-dimensional real coordinates. Using the
complex notation, some calculations become more easy and some deep mathematical theorems can
be derived. Here I show that 
the Taylor-expansion for the functions $g_1$ and $g_2$ as defined
in Eq.~(\ref{g1g2eq}) can be easily obtained using complex notation as follows.

With $z = x_1 + \mathrm{i}\,x_2$, one can write these functions as
\begin{equation}
g_1(x_1,x_2) = \Re f(\overline{z})\,,\quad g_2(x_1,x_2) = \Im f(\overline{z})\,,
\end{equation}
where
\begin{equation}
f(\overline{z}) = \frac{z}{|z|^2} = \frac{1}{\overline{z}}
\end{equation}
is an analytical function of $\overline{z}$.
This function can be expanded into a power series around a point $\overline{z_0}$ given by
\begin{equation}
f(\overline{z_0}+\overline{z}) = \sum_{n=0}^{\infty} \frac{f^{(n)}(\overline{z_0})}{n!}\,
{\overline{z}^n}\,,
\end{equation}
where $f^{(n)}$ denotes the $n$-th derivative with respect to $\overline{z}$,
which is given by
\begin{equation}
f^{(n)}(\overline{z_0}) = (-1)^n\,\frac{n!}{\overline{z_0}^{n+1}}\,.
\end{equation}
This yields for the power series
\begin{equation}
f(\overline{z_0}+\overline{z}) = \sum_{n=0}^{\infty} (-1)^n\,\frac{{\overline z}^n}{{\overline{z_0}^{n+1}}}\,.
\end{equation}
For $z_0 = (x_0,0) = \overline{z_0}$, one obtains
\begin{eqnarray}
f(x_0+x_1 - \mathrm{i}\,x_2) & = & \sum_{n=0}^{\infty} \frac{(-1)^n}{x_0^{n+1}}\,(x_1 - \mathrm{i}\,x_2)^n
\nonumber \\
& & \!\!\!\!\!\!\!\!\!\!\!\!
=  \sum_{n=0}^{\infty} \frac{(-1)^n}{x_0^{n+1}} \sum_{l=0}^{n} {\textstyle{n \choose l}}\,
x_1^{n-l} x_2^l\,(-1)^l\,\mathrm{i}^l\,.
\label{ser1}
\end{eqnarray}
Separating this into real and imaginary part yields the expressions given by Eq.~(\ref{gexp}).
Note that even values of $l$ contribute to $g_1$ while odd values of $l$ contribute
to $g_2$.

To investigate for which values of $(x_1,x_2)$ this series converges, let us first investigate the case
$x_2 = 0$, in which the series reduces to
\begin{equation}
f(x_0+x_1) = \sum_{n=0}^{\infty} \frac{(-1)^n}{x_0^{n+1}}\,x_1^n\,.
\label{ser2}
\end{equation}
This series is a geometrical series
\begin{equation}
f(x_0+x_1) = f_0\,\sum_{n=0}^{\infty} r^n
\end{equation}
with 
\begin{equation}
f_0 = 1/x_0\,,\quad r = -x_1/x_0\,.
\end{equation}
Since the geometrical series converges for $|r| < 1$, the series for $f$ 
in the case $x_2 = 0$ (Eq.~(\ref{ser2})) converges
for $|x_1| < |x_0|$.
Since a complex power series converges inside a 'convergence radius' and diverges outside, the series
for $f$ (Eq.~(\ref{ser1})) converges for $|z| < |x_0|$, i.e. for $\sqrt{x_1^2 + x_2^2} < |x_0|$.  

\end{appendix}

\end{document}